\documentclass[footheight=65pt]{scrartcl}
\pdfoutput=1

\usepackage{hyperxmp}
\usepackage[unicode]{hyperref}
\usepackage{graphicx}
\usepackage{booktabs}
\usepackage{xspace}
\usepackage{xcolor}
\usepackage[algoruled,linesnumbered,noend]{algorithm2e}
\usepackage{amsthm}
\usepackage{multirow}
\usepackage{authblk}
\newtheorem{example}{Example}[section]
\newtheorem{theorem}{Theorem}
\newtheorem{remark}{Remark}
\newtheorem{definition}{Definition}
\newtheorem{lemma}[theorem]{Lemma}
\newtheorem{proposition}{Proposition}

\usepackage{microtype}

\newcommand{\al}{alph}

\newcommand{\CFL}{\ensuremath{\mathsf{CFL}}\xspace}
\newcommand{\ICFL}{\ensuremath{\mathsf{ICFL}}\xspace}
\newcommand{\CFLICFL}{\ensuremath{\mathsf{CFL\_ICFL}}\xspace}
\newcommand{\ICFLCFL}{\ensuremath{\mathsf{ICFL\_CFL}}\xspace}

\newcommand{\comb}{\ensuremath{^{d}}}
\newcommand{\fact}{\ensuremath{F}}

\newcommand{\fingerprint}{\ensuremath{\mathcal{L}}}

\newcommand{\intseq}{\mathcal{I}}
\newcommand{\ie}{\textit{i.e.},\xspace}

\hypersetup{%
  pdftitle={Numeric Lyndon-based feature embedding of sequencing reads for machine learning approaches},
  pdfauthor={Paola Bonizzoni, Matteo Costantini, Clelia De Felice, Alessia Petescia, Yuri Pirola, Marco Previtali, Raffaella Rizzi, Jens Stoye, Rocco Zaccagnino, Rosalba Zizza},
  colorlinks = true,
  urlcolor = red!50!black,
  citecolor = blue!90!black,
  linkcolor = red!50!black,
  pdfpagemode = UseNone,
}

\usepackage{scrlayer-scrpage}

\usepackage[
    type={CC},
    modifier={by-nc-nd},
    version={4.0},
    imagewidth={4em},
    imageposition={left},
    imagedistance={1em},
]{doclicense}

\makeatletter
\renewcommand{\doclicenseLongText}{%
  \footnotesize\textcopyright{} 2022. \doclicense@lang@thisDoc\space
  \href{\doclicenseURL}{\doclicenseLongType\space\enquote{\doclicense@longName}}%
  \doclicense@lang@word@license.\xspace%
}
\makeatother
\ifoot[\doclicenseThis]{}
\begin{document}

\title{Numeric Lyndon-based feature embedding of sequencing reads for machine learning approaches}

\author[1]{Paola Bonizzoni}
\author[1]{Matteo Costantini}
\author[2]{Clelia De Felice}
\author[1]{Alessia Petescia}
\author[1]{Yuri Pirola}
\author[1]{Marco Previtali}
\author[1]{Raffaella Rizzi}
\author[3]{Jens Stoye}
\author[2]{Rocco Zaccagnino}
\author[2]{Rosalba Zizza}
\affil[1]{Dip.~di Informatica, Sistemistica e Comunicazione, University of Milano-Bicocca, Milan, Italy}
\affil[2]{Dip.~di Informatica, University of Salerno, Salerno, Italy}
\affil[3]{Faculty of Technology and Center for Biotechnology (CeBiTec), University of Bielefeld, Bielefeld, Germany}

\date{}

\maketitle

\begin{abstract}
 Feature embedding methods have been proposed in literature to represent sequences as numeric vectors to be used in some  bioinformatics investigations, such as family classification and protein structure prediction.

Recent theoretical results showed that the well-known Lyndon factorization preserves common factors in overlapping strings~\cite{tcs2020}.
Surprisingly, the \textit{fingerprint} of a sequencing read, which is the sequence of lengths of consecutive  factors in variants of the Lyndon factorization of the read, is effective in preserving sequence similarities, suggesting it as basis for the definition of novels representations of sequencing reads.

We propose a novel feature embedding method for Next-Generation Sequencing (NGS) data using the notion of \emph{fingerprint}. We provide a theoretical and experimental framework to estimate the behaviour of fingerprints and of the $k$-mers extracted from it, called $k$-fingers, as possible feature embeddings for sequencing reads. As a case study to assess the effectiveness of such embeddings, we use fingerprints to represent  RNA-Seq reads and to assign them to the most likely gene from which they were originated as fragments of transcripts of the gene.

We provide an implementation of the proposed method in the tool \textit{lyn2vec}, which produces Lyndon-based feature embeddings of sequencing reads.

\end{abstract}

\section{Introduction}

With the massive growth of many types of data in the Big Data era, Data Mining and Data Analytics are fundamental instruments for discovering interesting patterns in such data.

In particular, mining \textit{sequence} data has attracted a lot of attention for two main reasons. First, sequential representation for data and events is common for many real-life applications.
Second, a number of applications can benefit from knowing  useful patterns from sequences, such as, for example, web access analysis, event prediction, pattern discovery, time-aware recommendation, DNA sequence detection~\cite{biosensors2002} and \textit{feature embedding}~\cite{kumar2011pattern}.
Feature embedding is a challenging task in sequence mining~\cite{GAN2020222}; its main goal is to provide a machine-interpretable representation for sequence data that may improve the performance of learning algorithms.

In the specific context of machine learning applications involving \textit{biological sequences}, feature embedding is often based on the  assumption that there exists a  conceptual analogy between the languages adopted by humans to communicate and the sophisticated languages used by biological organisms to convey information within  cells. Most of the approaches proposed in the literature~\cite{asgari2015continuous,kimothi2016distributed}, indeed, adopt existing methods in Natural Language Processing (NLP), such as \texttt{word2vec}~\cite{mikolov2013efficient}, with the goal of  discovering functions encoded by biological sequences~\cite{yandell2002genomics,searls2002language}.

Typically, such methods involve fixed-length overlapping \emph{n-grams}~\cite{srinivasan2013mining,vries2008subfamily} which are also common in various techniques in Bioinformatics for analyzing sequences. However,  n-grams are not directly used in feature extraction, but for training an embedding model that is then used for feature extraction. We remark that most of the existing methods extract only  short-term patterns, otherwise  they exhibit a significant increase in the computational time, which is a main  limitation in addition to those  discussed in Section \ref{sec:related}.

\subsection{Contributions of this work}
In this article, we focus on a novel approach for the feature embedding of
sequencing reads. Differently from the works proposed in the
literature which essentially consist of elaborate
applications of NLP techniques, we propose a theoretical investigation of combinatorial properties that would guarantee compact embedded representations of the sequences while preserving similarities.
In this section, we provide a discussion on the motivations; then, we give an overall description of the proposed method, and finally we present our main contributions.
\paragraph{Motivations.}
The main question addressed in this paper is whether there exists a ``similarity signature'' that can be: \textit{(i)} easily detected while reading the sequence, and \textit{(ii)} used to define an effective feature embedding method. We answer to this question by exploiting one of the most well-known factorizations in combinatorics on words: the \textit{Lyndon factorization}~\cite{10.2307/1970044,lothaire1997combinatorics}.
Such a factorization has some main desired properties:
\textit{(i)} it is unique for a word, \textit{(ii)} it can be computed in linear time, and more recently, it has been proved in~\cite{tcs2020,lata2020} that, under some conditions, the sequence of Lyndon factors of a word shares consecutive common factors with the Lyndon factorization of a superstring of the word itself.
The notion of Lyndon word is not novel in the field of Bioinformatics, since it was used to locate short motifs~\cite{delgrange2004star} and more recently it was explored in the development of the bijective Burrows-Wheeler Transforms~\cite{inplace-bijective-bwt2020}.

Surprisingly, we discovered that the sequence of lengths of the factors in Lyndon factorizations is enough to capture sequence similarity, thus allowing the definition of a word signature, called \textit{fingerprint}.

\paragraph{Proposed method.} In this paper we propose a novel feature embedding method for sequences generated by sequencing, called \emph{reads}, that exploits the fingerprint of such sequences to produce their embedded representations.
To the best of our knowledge, it represents the first attempt to build a feature representation that is based on theoretical combinatorial properties proved to capture sequence similarities and also suitable for machine learning approaches.

\paragraph{Our contributions.}
The main contributions can be summarized as follows:
\begin{enumerate}
    \item We have implemented the proposed method in the tool \texttt{lyn2vec}, which produces Lyndon-based feature embeddings of sequencing reads.

    \item Unlike NLP-based embedding methods, \texttt{lyn2vec} does not require any previous training on a text corpus, but it is based on combinatorial properties that capture sequence similarities. In Sections~\ref{ssec:theo-investigation} and~\ref{sec:evaluation}, we investigate properties of the embedded representations with respect to some parameters, such as the specific variant of the Lyndon factorization used to compute the fingerprint and the value of $k$ used to extract the $k$-fingers from the fingerprint.
    \item The computational complexity of \texttt{lyn2vec} is related to the running time of the algorithms for computing  the Lyndon factorization; as described in Section~\ref{ssec:theo-investigation}, these algorithms run in linear time in the length of the sequence, hence, the time complexity of the \texttt{lyn2vec} is linear in the total length of all sequences.

    \item We introduce the theoretical notion of  \emph{collision rate} to investigate some limitations of the use of Lyndon-based representations and we address the problem of how the collision phenomenon may be affected by the lexicographic ordering of the alphabet.

    \item As a proof of concept of the possible uses in machine learning tasks of the proposed embedding representations, we have evaluated the effectiveness of such representations in classifying reads to the most likely genes from which they originated: we obtained satisfactory precision and recall even for the case of chimeric reads, i.e., reads originating from two fused genes.
\end{enumerate}

This paper is organized as follows. In Section~\ref{sec:related} we discuss the main contributions in the literature and the differences with our work. In~Section~\ref{sec:lynd-representations} we present the properties of various notions of Lyndon-based factorization, including one inspired by the the double-stranded nature of the genome. In Section~\ref{sec:evaluation}, we propose some experiments to evaluate the effectiveness of the representations produced by \texttt{lyn2vec}.  Finally, in Section~\ref{sec:conclusion} we discuss the main results and the future work.

\section{Related work}
\label{sec:related}
The past years have seen impressive advances in sequence mining.
State-of-the-art sequence mining methods could be categorized as follows: \textit{(i)} sequence alignment~\cite{stoye1997dca}, \textit{(ii)} string kernels~\cite{leslie2004mismatch}, \textit{(iii)} time-series classification~\cite{bagnall2015time,hills2014classification}, and \textit{(iv)} pattern discovery~\cite{didier2012variable,zaki2001spade}.
In the following, we will focus on the feature embedding of \textit{sequencing reads}.

Recent results in NLP~\cite{mikolov2013efficient} identify  terms with a similar linguistic context by exploiting  \textit{word embedding}. In this approach, known as \texttt{word2vec}, words or phrases are mapped to vectors of real numbers in a low-dimensional space. By training a neural network over a large
text corpus, words with similar linguistic context correspond to vectors that are close points in the Euclidean space.

In~\cite{asgari2015continuous}
the \texttt{word2vec} framework is applied to extract features from biological sequences. The embeddings generated for general biological sequences are named \texttt{BioVec} for genomic sequence and \texttt{ProtVec} for proteins.
The tool \texttt{seq2vec}~\cite{kimothi2016distributed} extends the idea proposed in~\cite{asgari2015continuous} by modeling a sequence as a sentence in a text corpus  while  $k$-mers derived from such a sequence are words in the sentence given as input to the embedding algorithm. In~\cite{fastDNA}, a novel model for fast classification of sequencing reads is proposed, which is named \texttt{fastDNA}. It is based on \texttt{FastText}~\cite{bojanowski2016enriching}, which is an extension of \texttt{word2vec}, where the main difference is that instead of using individual words to train the Neural Network, words are broken into several n-grams used to train the network.

In~\cite{DNABERT}, another well-known NLP model, the Bidirectional Encoder Representations from Transformers (BERT), is adapted to model DNA general embeddings. The result is \texttt{DNABERT}, a novel pre-trained bidirectional representation encoder for DNA-language.

These sequence embedding methods   essentially consist of elaborate applications of NLP techniques, in which an embedding model is first trained on a large text corpus, and then used to transform biological sequences into numeric vectors to be used by learning algorithms.
Therefore, the computational space  and time required to embed a sequence dataset is a  critical issue  of these embedding methods.
Instead, in our approach we change perspective: we investigate combinatorial properties that would guarantee compact embedding representations able to preserve similarities. As a result, we define novel embedding representations which can be computed, for each sequence, in linear time with respect to the length of the sequence itself, without requiring any previous training. Furthermore, the ability to capture sequence similarities can be controlled by tuning some parameters that are determined by the specific factorization used.

\section{ Variants of the Lyndon factorization and \texttt{lyn2vec}}
\label{sec:lynd-representations}

In this section, we present the well-known \textit{Lyndon factorization}
and some variants that have been recently introduced. The Lyndon factorization will be denoted by the acronym
$\CFL$, used for the first time in~\cite{duval1983factorizing}
and built using the initial letters of the surnames of the authors introducing this factorization~\cite{10.2307/1970044}.
Then, the related notion of \textit{fingerprint} and the \texttt{lyn2vec} method will be presented. Observe that \texttt{lyn2vec} is the name we give to the different Lyndon-based sequence embeddings: it is indeed a tool implementing various representations.

\subsection{Variants of the Lyndon factorization and overlapping strings}
\label{ssec:theo-investigation}

\paragraph{Basics.} Let $\Sigma$ be a finite alphabet and let $s = a_1 \cdots a_n$ be a
{\em string} over $\Sigma$, \ie a sequence of
$n$ characters of $\Sigma$; the \emph{length} $n$ will be denoted by $|s|$.
The character of $s$ at position $i$ (that is, $a_i$) is denoted by $s[i]$.
The \textit{substring} (or {\em factor})
of $s$ from position $i$ to position $j$ is denoted by $s[i:j]$.
When $i=1$ (resp. $j=|s|$) the factor is named
\textit{prefix} (resp. \emph{suffix}) of $s$,
also denoted by $s[:j]$ (resp. $s[i:]$).
In addition, if $j \ne |s|$ (resp. $i \ne 1$),
the prefix (resp. suffix) of $s$ is \emph{proper}.

In the following, $\Sigma$ is
supposed to be totally ordered with respect to the lexicographic order, denoted by $<$.
We classically extend this notion on $\Sigma^*$,
by defining  $s \preceq v$
if $s$ is a prefix
of $v$ or $s = xay$, $v = xbz$, with $a < b$
and $x,y,z \in \Sigma^*$.
Moreover, $s \prec v$ if $s \preceq v$
and $s \not = v$.
Symmetrically, can define $v \succeq s$ (resp. $v \succ s$) if $s \preceq v$
(resp. $s \prec v$)~\cite{bonizzoni2018inverse}.
For two nonempty strings $s,v$, we write $s \ll v$ if $s \prec v$ and $s$ is not a proper prefix of $v$~\cite{Bannai15}.

Finally, due to the particular context considered for assessing the proposed embedding representations (Section~\ref{sec:evaluation}),
the definition of \emph{reverse and complement} of a string $s$ over the DNA alphabet $\{A,C,G,T\}$ is needed.
This is a typical notion in Bioinformatics originating from the double-stranded nature of reads sequenced from the genome.
Precisely, the {\em reverse} of $s$ is
obtained by reading $s$ from right
to left and the (Watson-Crick) \emph{complement} of a DNA symbol is the operation transforming $A$ into $T$ (or vice versa) and  $C$ into a $G$ (or vice versa).
Thus, the \emph{reverse and complement} of $s$
is an application transforming $s$ in $\overline{s}$, where $\overline{s}$ is obtained by taking the reverse $s^r$ of $s$ and by replacing each symbol in $s^r$ with its complement.

\paragraph{Definitions.} We recall the notion of a \emph{factorization} and
introduce the related notion of \emph{fingerprint}. These are the main
ingredients we use to capture the overlap between two reads.
A \emph{factorization} of a string $s$
is a sequence $\langle f_1, \ldots, f_n \rangle$ of strings such
that $s=f_1 \cdots f_n$.
We mainly consider factorizations which can be derived by using algorithms.
The notation $\fact(s) = \langle f_1, f_2, \dots, f_n \rangle$ means that the factorization of $s$ is obtained  by using a factorization algorithm
$\fact$.
The \emph{fingerprint} of $s$ with respect
to $\fact(s)$ is the sequence $\fingerprint(s)$ of the lengths of the factors in $\fact(s)$, that is,
$\fingerprint(s) = \langle |f_1|, |f_2|, \dots, |f_n| \rangle$.

Given a fingerprint
$\fingerprint(s) = \langle l_1, l_2, \ldots, l_n \rangle$ and an integer $k$ ($1 \leq k \leq n$)
a \emph{$k$-finger} is any subsequence
$\langle l_i, l_{i+1}, \ldots, l_{i+k-1} \rangle$ of $k$ consecutive lengths, that is, a $k$-mer of $\fingerprint(s)$.
Let $\fingerprint(s)$ be a fingerprint
of $s$ with respect to $\fact(s) = \langle f_1, f_2, \dots, f_n \rangle$.
Given a $k$-finger
$\langle l_1, \ldots, l_{k} \rangle$, a {\em supporting string} of it is
any subsequence $\langle f_j, \dots, f_{j+k-1} \rangle$
of $k$ factors of $\fact(s)$ such that $|f_j|=l_1, \ldots, |f_{j+k-1}|=l_k$. A $k$-finger may have more than one supporting string, as the following
Example~\ref{esempio_k} shows.
The side effect of this fact will be discussed in Section~\ref{ssec:collisions}.

\begin{example} \label{esempio_k_2}
Let us consider the factorization $F(s) = \langle GC, \mathit{ATC}, \mathit{ACCTCT}, CT,$ $\mathit{ACAG}, \mathit{TAT}, A \rangle$. Then, the fingerprint is $\mathcal{L}(s) = \langle 2,3,6,2,4,3,1 \rangle$ and $\langle 2,4,3 \rangle$ is a $k$-finger for $k=3$, whose \emph{supporting string} is $\mathit{CTACAGTAT}$, given by the concatenation of the fourth, fifth and sixth factors $CT$, $\mathit{ACAG}$ and $\mathit{TAT}$.
\end{example}

\begin{example} \label{esempio_k}
Let us consider the factorization $\fact(s)=\langle b,b,ababb,a,a \rangle$.
Then, $\mathcal{L}(s) = \langle 1,1,5,1,1 \rangle$ and $\langle 1,1 \rangle$ is a $k$-finger for $k=2$.
Observe that  $bb$ and $aa$ are both supporting
strings for the $2$-finger  $\langle 1,1 \rangle$.
\end{example}

Fingerprints used for capturing overlaps, are based on factorizations in \emph{Lyndon Words}~\cite{lyndon1954burnside, berstel2007origins}.
A nonempty string $s$ is a \textit{Lyndon word} if and only if it is strictly smaller than any of its nonempty proper suffixes.
For example, $s=aabbab$ over alphabet $\{a, b\}$, $a<b$, is a Lyndon word, whereas string $s'=abaabb$ is not a Lyndon word, since the suffix $aabb$ is smaller than $s'$.
It is well-known that
any nonempty string $s$ has a unique factorization $\langle f_1, f_2, \dots, f_n \rangle$
such that
$f_1 \succeq f_2 \succeq \dots \succeq f_n$ and each factor $f_i$ is a
Lyndon word. Such a factorization is called
the \emph{Lyndon factorization} and can be computed in linear time and constant space by the Duval algorithm~\cite{duval1983factorizing}. As usual, $\CFL(s)$ denotes the Lyndon factorization of $s$
\cite{10.2307/1970044,lothaire1997combinatorics}.

\begin{example} \label{esempio_CFL}

The factorization $\langle b,b,ababb,a,a \rangle$ (of Example~\ref{esempio_k}) is a \CFL factorization and the \CFL factorization of string
$s'=aaabbbabab$  is trivially $\langle aaabbbabab \rangle$, since $s'$ is itself a Lyndon word.
\end{example}

\paragraph{Conservation Property.}

We now report a crucial property of $\CFL$, provided in~\cite{tcs2020}.
To this aim, we need to recall the definition of \textit{simple factor} (or simple substring) given in~\cite{lata2020}.
A substring $x$ occurring in a string $s$ is \textit{simple} with respect to a factorization $\fact(s) = \langle f_1, f_2, \dots, f_n \rangle$ if, for each occurrence of $x$ in $s$, there is an index $j$, $1 \leq j \leq n$ such that $x$ is a substring of $f_j$.
Informally, every occurrence of of a \emph{simple substring} $x$ needs to be within some factor $f_j$.
We say that $x$ is a {\em simple prefix}
(resp. {\em simple suffix}) of $s$ if $x$
is a proper prefix (resp. suffix) of $f_1$ (resp.
$f_n$).

Let $w =xz$ and $w'= zy$ be two substrings of a string $s = xzy$,
\ie $w$ and $w'$ share a common overlap $z$ in $s$ let us suppose that $z$ is a non-simple substring
with respect to $\CFL(s)$
(as a suffix of $w$ and as a prefix of $w'$).
As a consequence of Lemma 13.2 in~\cite{tcs2020}, $\CFL(xz)$ and $\CFL(zy)$ may share common Lyndon factors with $\CFL(xzy)$.
Moreover, some of these factors may be in $z$. More precisely, given $w =xz$ and $w'= zy$,
let us assume that $\CFL(w) = \langle h_1, h_2, \dots, h_n \rangle$ and $\CFL(w') = \langle g_1, g_2, \dots, g_m \rangle$.
If $z$ is both a non-simple suffix of $w$ and a non-simple prefix of $w'$, then
there are two indexes $i,j$, with
$1 \leq i < n$, $1 < j \leq m$, such that
$z = h''_i h_{i+1} \cdots h_{n} =
g_1 \cdots g_{j-1}g'_j$,
where $h''_i$ is a suffix of $h_i$ and
$g'_j$ is a prefix of $g_j$.
Let
$\CFL(h''_i) = \langle m_1, \ldots , m_r \rangle$ and
$\CFL(g'_j) = \langle v_1, \ldots , v_t \rangle$. Hence, as a consequence of the above result, we have
$\CFL(z) = \langle m_1, \ldots , m_r, h_{i+1}, \ldots , h_{n} \rangle =
\langle g_1, \ldots ,g_{j-1}, v_1, \ldots , v_t \rangle$.

\begin{figure}
	\includegraphics[width=0.99\linewidth]{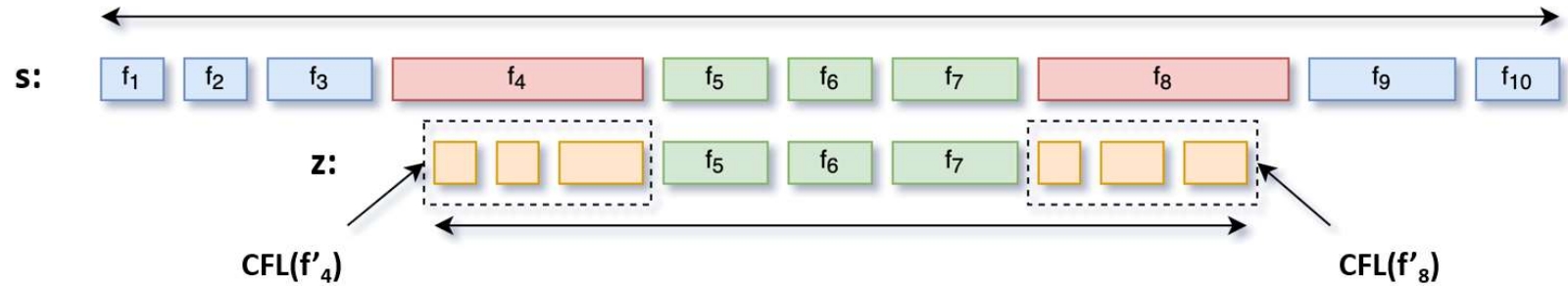}
\caption{Conservation Property: two $\CFL$ factorizations are schematically depicted for a string $s$, that is, $\CFL(s) = \langle f_1, f_2, \dots, f_{10} \rangle$ (above) and a substring $z$, that is, $\CFL(z) = \langle \CFL(f'_4), f_{5}, f_{6}, f_{7}, \CFL(f_{8}^{'}) \rangle$ (below) covering a suffix of factor $f_4$ (of $s$), the whole factors $f_5$, $f_6$ and $f_7$ and finally a prefix of $f_8$.
The green factors $f_5$, $f_6$ and $f_7$ in the middle are conserved between the two factorizations.}
\label{fig:conservation}
\end{figure}

Furthermore,
let us suppose that $\CFL(s)=\langle f_1, f_2, \dots, f_n \rangle$. Let
$z=f'_l f_{l+1} \cdots f_{t} f_{t+1}^{'}$
be a non-simple factor with respect to $\CFL(s)$,
for some indexes $l, t$
with $1 \leq l <n $, $1 < t < n$, and let us assume
$f_l=f''_lf'_l$,
$f_{t+1}=f'_{t+1}f''_{t+1}$.
By abuse of notation,
in view of the above discussion on Lemma 13.2,
we have that
$\CFL(z)= \langle \CFL(f''_l), f_{l+1}, \ldots f_{t}, \CFL(f_{t+1}^{'}) \rangle$
(see Figure \ref{fig:conservation}).
The same argument applies if we consider
strings $xz$ and $zy$ and their Lyndon factorizations. Let us assume that $z$ is a non-simple factor with respect to both $\CFL(xz)$ and $\CFL(zy)$.
Thus, $\CFL(z)$ shares factors with $\CFL(xz)$
and with $\CFL(zy)$. Since $\CFL(z)$ is unique, then there will exist factors that are clearly in common between $\CFL(xz)$ and $\CFL(zy)$.
It follows that two overlapping strings $w$ and $w'$ may share consecutive common Lyndon factors in their Lyndon factorizations. Thus,
the fingerprints of $w$ and $w'$ will share consecutive integers.

\begin{remark}\label{oss:z}
We explicitly point out that
the hypothesis that $z$ is not a simple factor with respect to $\CFL(s)$ cannot be dropped. Indeed, consider $w=xz=bbaababaa$,
$w'=zy=ababaabbbbbb$, where $z=ababaa$.
We have
$\CFL(w)=\CFL(xz)=\langle b,b,aabab,a,a \rangle$
and
$\CFL(w')=\CFL(zy)=\langle ab,ab,aabbbbbb\rangle$.
Even though $z$ is an overlapping factor,
the two factorizations do not share consecutive
factors. Observe that this is not a counterexample, as
$\CFL(xzy)=\langle b,b,aababaabbbbbb \rangle$,
\ie $z$ is non-simple.
\end{remark}

Such an interesting property suggests the possibility of using directly $k$-fingers as features. Indeed, as we will see in Section~\ref{sec:evaluation},
to assess such an intuition, we also propose an
approach in which we use $k$-fingers for classifying sequencing reads.

\paragraph{Lyndon-based factorizations.}
As a consequence of the above discussion, two main properties can be observed: \textit{(i)} the fingerprint ability of preserving similarities,
\textit{(ii)} the possibility of tuning the $k$ value to control the number of  factors included in overlapping substrings. The latter property suggests a further investigation on variants of the Lyndon factorization, we call \textit{Lyndon-based} factorizations. Specifically, we introduce two types of Lyndon-based factorization: the \textit{single-stranded} factorization and the \textit{double-stranded} factorization, for dealing with sequencing reads derived from a unique genome strand or from both strands of a genome, respectively.

At this point, we recall the notion of \textit{inverse Lyndon word}
given in~\cite{bonizzoni2018inverse}: a nonempty
string $s$ is an inverse Lyndon word if each proper suffix is strictly smaller than $s$.
For instance, $a, b, aaaaa, bbba, baaab, bbaba$ are inverse Lyndon words over the alphabet $\{a, b\}$, such that $a < b$.

A factorization $\fact(s) = \langle f_1, f_2, \dots, f_n \rangle$ of a
string $s$ is an \emph{inverse Lyndon factorization} if $f_j$ is an
inverse Lyndon word for $1 \le j \le n$ and $f_1 \ll f_2 \ll \dots \ll f_n$.
In~\cite{bonizzoni2018inverse}, a linear time algorithm is proposed to produce a special inverse Lyndon factorization which is unique for the string and is
called \emph{Canonical inverse Lyndon factorization} or \textit{inverse $\CFL$}, referred in the following by $\ICFL$.
As well as $\CFL$, $\ICFL$ guarantees uniqueness and linear time computation. In addition, a factor of $\ICFL$ cannot be a prefix of the next one by definition, thus $\ICFL$ is less prone to split a string in two different factors. Interestingly, as proved in~\cite{bonizzoni2018inverse},
$\ICFL$ allows to split any Lyndon word, thus allowing to further factorize long Lyndon factors of Lyndon factorizations (as well as $\CFL$ allows to split any inverse Lyndon word).

\begin{example} \label{esempio_ICFL}
Let us consider $s=bbababbaa$ and $s'=aaabbbabab$ of Example \ref{esempio_CFL}.
We have that $\ICFL(s)=\langle s \rangle$, since $s$ is an inverse Lyndon word, and $\ICFL(s')=\langle aaa,bbbabab\rangle$.
\end{example}

For an extended discussion of the theoretical background of these factorizations, we refer to~\cite{bonizzoni2018inverse}.

Given a threshold $T$, we call $\CFLICFL_T$ the factorization obtained by first computing the $\CFL$ factorization, and next by performing a further factorization of the $\CFL$ factors longer than $T$ by means of the $\ICFL$ algorithm. The subscript $T$ will be omitted when it is clear from the context.
In more detail, given
$\CFL(s) = \langle f_1, f_2, \dots, f_n \rangle$, we obtain $\CFLICFL$ by replacing each $f_i$ longer than $T$ with $\ICFL(f_i)$.
A symmetrical definition can also be obtained by first applying $\ICFL$ and next $\CFL$, thus defining the $\ICFLCFL$ factorization. The examples below show the $\CFLICFL$ and $\ICFLCFL$ factorizations over a simple string.
Note that $\CFLICFL$ has several advantages over $\CFL$ or $\ICFL$ alone.
Indeed, the $\ICFL$ factorization, applied to long factors of a $\CFL$ factorization (or vice versa), concurs to increase the number of factors.

\begin{example} \label{esempio_CFL_ICFL}
Let $s=dabadabdabdadac$ be a string ($a<b<c<d$).
We have $\CFL(s)=\langle d,abadabdabdadac \rangle$ and $\ICFL(s)=\langle daba,dabdab,dadac \rangle$.
Assuming $T=1$, we can factorize the second factor of $\CFL(s)$ by applying $\ICFL$, thus obtaining $\ICFL(abadabdabdadac)= \langle a,ba,dabdab,dadac \rangle$ and $\CFLICFL(s)=\langle d,a,ba,dabdab,dadac \rangle$.
\end{example}

\begin{example} \label{esempio_ICFL_CFL}
Let $s=adbadbadba$ be a string ($a<b<c<d$).
We have $\CFL(s)=\langle adb,adb,adb,a \rangle$ and $\ICFL(s)=\langle a, \mathit{dbadbadbadba} \rangle$.
Assuming $T$ equal to $1$, we obtain
$\ICFLCFL(s)=\langle a,d,b,adb,adb,a \rangle$.
\end{example}

Example~\ref{esempio_CFL_ICFL} shows that $\ICFL$ provides better results than $\CFL$ in terms of factor length distribution, whereas Example~\ref{esempio_ICFL_CFL}, shows that $\CFL$ is better than $\ICFL$.
It is not difficult to see that the Conservation Property holds even for $\CFLICFL$. Indeed, if $\CFL(xz)$ and $\CFL(zy)$ share a Lyndon factor $f_{h+1}$
and $f_{h+1}$ is split by $\CFLICFL(xz)$, then the same factor $f_{h+1}$ is split in the same
way by $\CFLICFL(zy)$, because in both cases we replace $f_{h+1}$ with $\ICFL(f_{h+1})$.
On the other hand, we do not know whether a conservation property holds for $\ICFL$ and $\ICFLCFL$, even though experiments reported in the paper suggest a positive answer.

\begin{figure}
	\includegraphics[width=0.99\linewidth]{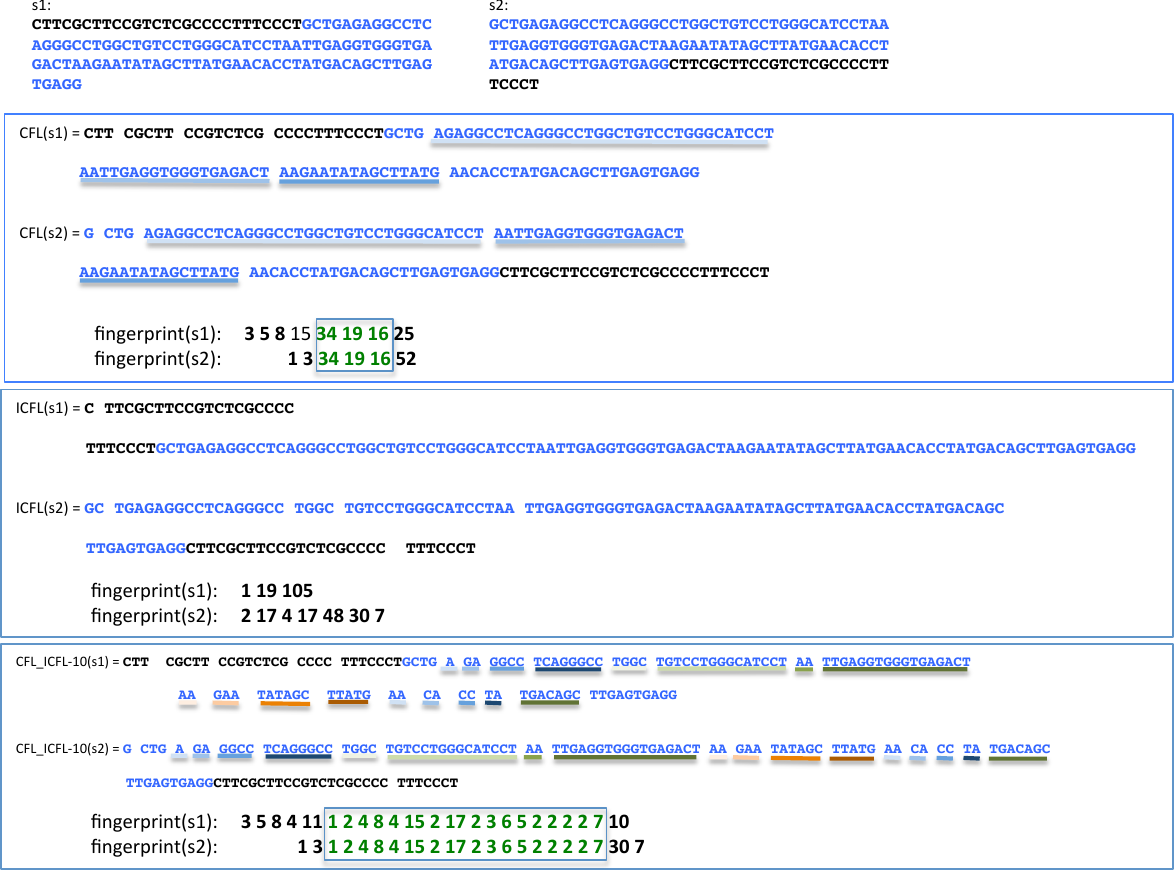}
	\caption{The factorizations $\CFL$, $\ICFL$ and $\CFLICFL_{10}$ are depicted for two overlapping 125-long reads $s1$ and $s2$ (the overlap is highlighted in blue) together with their fingerprints. The common factors are underlined. $\CFL$ produces the common $3$-finger  $\langle 34,19,16 \rangle$, whereas there are no common factors for $\ICFL$ (observe the long factors at the end of each read). $\CFLICFL_{10}$, obtained by applying first $\CFL$ and then $\ICFL$ to factors longer than $10$, yields $17$ consecutive common factors between the reads, and a common $17$-finger in the fingerprints.}
	\label{fig:read-factorizations}
\end{figure}

Figure~\ref{fig:read-factorizations} shows an example of $\CFL$, $\ICFL$ and $\CFLICFL$ for two 125-long overlapping reads.

\paragraph{Double-stranded factorization.}
We are interested in signatures able to highlight the common regions between two overlapping reads originating from opposite strands of the genome.
To this aim, given a Lyndon-based factorization $\fact$ (Section~\ref{ssec:theo-investigation}), we introduce the definition of \emph{double-stranded} factorization $\fact\comb$ built on on a basic algorithm $\fact$, which has the fundamental property  stated by the following Definition~\ref{prop:comb}.
Indeed, we seek for a factorization algorithm $F\comb$ such that the factorization $\fact\comb(\overline{s})$ of the reverse and complement $\overline{s}$ of a string $s$ is equal to the reversed list of the reverse and complement of the factors of $\fact(s)$.
We show (Theorem \ref{prop:comb}) that there exists an algorithm for constructing such a factorization, by appropriately combining the fingerprints obtained from $\fact(s)$ and $\fact(\overline{s})$.

\begin{definition} \label{prop:comb}
Given a factorization algorithm $F$, a \emph{double-stranded} factorization of a string $s$ is a factorization $\fact\comb(s) = \langle f_1, f_2, \dots, f_n \rangle$ such that, $\fact\comb(\overline{s}) = \langle \overline{f}_n, \overline{f}_{n-1}, \dots, \overline{f}_1 \rangle$, where $\overline{s}$ is the reverse and complement of $s$ and $\overline{f}_{i}$ is the reverse and complement of factor $f_i$, $1 \leq i \leq n$.
\end{definition}

It follows that, if $\fingerprint(s)=\langle l_1, l_2, \dots, l_n \rangle$ is the fingerprint of $s$ with respect to $\fact\comb$, then $\fingerprint(\overline{s}) = \langle l_n, l_{n-1}, \dots, l_1 \rangle$ will be the fingerprint of $\overline{s}$ with respect to
$\fact\comb$.
Note that $l_i = |f_i|$ and $\fingerprint(\overline{s})=\langle l'_1, l'_2, \ldots, l'_n \rangle$ where $l'_i = |\overline{f}_{n-i+1}|$.
In particular,
given a $k$-finger $\langle l_i, l_{i+1}, \ldots, l_{i+k-1} \rangle$ of $\fingerprint(s)$, we refer to the $k$-finger %
$\langle l'_{n-i-k+2}, l'_{n-i-k+3}, \ldots, l'_{n-i+1} \rangle$
of $\fingerprint(\overline{s})$ as its \emph{counterpart}, since they are clearly supported by the same string except for a \emph{reverse and complement} operation. In addition, note that any $k$-finger of $\fingerprint(s)$ is the reverse of its counterpart in $\fingerprint(\overline{s})$.
For this reason, in order to detect the common regions among a set of reads originating from both strands of the genome, it is necessary to perform a ``normalization'' operation on the extracted $k$-fingers. They are considered as strings over the alphabet of the positive integers and, for each extracted $k$-finger, the smallest (lexicographically) sequence between the $k$-finger and its reverse is retained.
For example, the ``normalization'' of $\langle 4,3,7,8,5 \rangle$ produces $\langle 4,3,7,8,5 \rangle$ (that is, the $k$-finger itself), whereas the ``normalization'' of $\langle 5,10,7,8,5 \rangle$ produces its reverse $\langle 5,8,7,10,5 \rangle$; indeed, $\langle 5,8,7,10,5 \rangle$ is smaller than $\langle 5,10,7,8,5 \rangle$, since the two sequences share the first element $5$ and the second element $8$ of the first sequence is smaller than the second element $10$ of the second one.

In the following, we %
describe the algorithm for constructing the double-stranded factorization for a string $s$ relying on a basic factorization algorithm $\fact$. Hence, we prove Theorem~\ref{prop:comb}.
To this purpose, we firstly introduce the notion of \textit{interval-sequence} of a fingerprint $\fingerprint(s)= \langle l_1, l_2, \dots, l_n \rangle$ as the sequence $\mathcal{I}(s) = \langle i_1, i_2, \dots, i_{n} \rangle$ such that  $i_j = l_1 + \cdots + l_j$ is the end position of factor $f_j$ on $s$.
Moreover, given a fingerprint $\fingerprint(s)$, we call \emph{reversed interval-sequence} $\mathcal{I}^r(s)$ the interval-sequence of the reverse of $\fingerprint(s)$.
For example, given  $\fingerprint(s) = \langle 1, 1, 7, 6 \rangle$, then $\mathcal{I}(s) = \langle 1, 2, 9, 15 \rangle$
is the interval-sequence and $\mathcal{I}^r(s) = \langle 6, 13, 14, 15 \rangle$ is the reversed interval-sequence.
More in general, the interval-sequence can be defined with respect to any sequence $\langle l_1, l_2, \dots, l_n \rangle$ of positive integers.

Moreover,
given a sequence $s$, any strictly increasing sequence $\langle i_1, i_2, \dots, i_{n} \rangle$ of positive integers, such that $i_n = |s|$,  {\em is associated} to a fingerprint $\langle l_1, l_2, \ldots, l_n \rangle$, where $l_1 = i_1$ and  $l_j = i_j - i_{j-1}$ for $1 < j \le n$, and induces on $s$ the factorization $\langle s[:i_1], s[i_1+1:i_2], \ldots, s[i_{n-1}+1:] \rangle = \langle s[:l_1], s[l_1+1:l_1+l_2], \ldots,
 s[l_1+ \cdots +l_{n-1}+1:] \rangle$.
At this point, given a factorization algorithm
$\fact$, the double-stranded factorization $\fact\comb$ introduced by Definition~\ref{prop:comb} can
be computed as stated by Theorem~\ref{th:prove-double},
where \emph{merging} two interval sequences simply means merging the two (sorted) lists of their integers, discarding repetitions.
Indeed, the elements of an interval-sequence have an intrinsic order, since each integer integer is a position on the string. Thus, the merger of two interval-sequences must preserve that order.

\begin{theorem}
\label{th:prove-double}
Let $\fingerprint(s)$ and $\fingerprint(\overline{s})$ be the fingerprints for a string $s$ and its reverse and complement $\overline{s}$ obtained by applying
 a basic factorization algorithm $\fact$. Let $\mathcal{I}(s)$ and $\mathcal{I}^r(\overline{s})$ be the interval-sequence associated to $\fingerprint(s)$ and the reversed interval-sequence associated to $\fingerprint(\overline{s})$, respectively.
Let us consider the increasing sequence ${\cal I}$ obtained by merging $\mathcal{I}(s)$ and $\mathcal{I}^r({\overline{s}})$,
and consider the fingerprint $\ell$ such that
$\cal I$ is associated to $\ell$. The factorization induced by $\ell$ is a \emph{double-stranded} factorization $\fact\comb(s)$ for $s$.
\end{theorem}

\begin{example}
	\label{ex:fact-comb}
	Let $s=\mathit{GGATCTCGCAGGCGG}$ be a string and $\overline{s} = \mathit{CCGCCTGCGAG}$ $\mathit{ATCC}$ its reverse and complement.
	By considering $\CFL$ as a basic factorization algorithm, we have $\CFL(s) = \langle G, G, \mathit{ATCTCGC}, \mathit{AGGCGG} \rangle$ and
	$\CFL(\overline{s}) = \langle \mathit{CCGCCTGCG},$ $\mathit{AGATCC} \rangle$.
	The fingerprint and the interval-sequence for $s$ are $\langle 1, 1, 7, 6 \rangle$ and $\langle 1, 2, 9, 15\rangle$ respectively, whereas the fingerprint and the reversed interval-sequence for $\overline{s}$ are $\langle 9, 6 \rangle$ and $\langle 6, 15 \rangle$, respectively. The increasing sequence $\langle 1, 2, 6, 9, 15 \rangle$, obtained by merging $\langle 1, 2, 9, 15\rangle$ and $\langle 6, 15 \rangle$ %
	is associated to the sequence $\langle 1, 1, 4, 3, 6 \rangle$, which induces on $s$ the double-stranded factorization $\CFL^d(s) = \langle G, G, ATCT, CGC, AGGCGG \rangle$. The double-stranded fingerprint of $s$ will be $\langle 1, 1, 4, 3, 6 \rangle$.

	Vice versa, the fingerprint and the
	interval-sequence for $\overline{s}$ are
	$\langle 9, 6 \rangle$ and $\langle 9, 15 \rangle$, while
	the fingerprint and the reversed interval-sequence for $s$ are $\langle 1, 1, 7, 6 \rangle$ and $\langle 6, 13, 14, 15 \rangle$.
	The increasing sequence $\langle 6, 9, 13, 14, 15 \rangle$, obtained by merging $\langle 9, 15 \rangle$ and $\langle 6, 13, 14, 15 \rangle$ is associated to the sequence $\langle 6, 3, 4, 1, 1 \rangle$, which induces on $\overline{s}$ the double-stranded factorization $\CFL^d(\overline{s})$ $=$ $\langle CCGCCT, GCG$, $AGAT, C, C \rangle$. The double-stranded fingerprint of $\overline{s}$ will be $\langle 6, 3, 4, 1, 1 \rangle$.
\end{example}

In the previous example, both factorizations $\CFL^d(s)$ and $\CFL^d(\overline{s})$ have five factors and the $i$-th factor of $\CFL^d(s)$ is the reverse and complement of  the $(5-i+1)$-th factor of $\CFL^d(\overline{s})$. For example, the fourth factor $CGC$ of $\CFL^d(s)$ is the reverse and complement of the "symmetrical" second factor $GCG$ of $\CFL^d(\overline{s})$. As a consequence, fingerprint $\langle 1, 1, 4, 3, 6 \rangle$ of $s$ is the reverse of fingerprint $\langle 6, 3, 4, 1, 1 \rangle$ of $\overline{s}$. The $2$-finger $\langle 1, 4 \rangle$ of $s$, supported by string $GATCT$, has its counterpart in the $2$-finger
$\langle 4, 1 \rangle$ of $\overline{s}$, which is supported by string $AGATC$: they are manifestly in a reverse and complement relation.

In order to prove that Theorem~\ref{prop:comb} allows to compute a double-stranded factorization as given by Definition \ref{prop:comb}, we need to introduce two technical lemmas.

\begin{lemma}
\label{prop:rev-comb1}
Let $\mathcal{L}_1 = \langle l_1, l_2, \dots, l_n \rangle$ and $\mathcal{L}_2  = \langle l'_1, l'_2, \dots, l'_n \rangle$, be two sequences with the same number $n$ of elements (positive integers).
Let $\intseq_1 = \langle i_1, i_2, \dots, i_n \rangle$ and $\intseq_2 = \langle i'_1, i_2^{'}, \dots, i_n^{'} \rangle$ be the respective interval-sequences.
Then, $\mathcal{L}_1$ is equal to the reverse of $\mathcal{L}_2$ iff $i_j = i'_n - i'_{n-j} \mbox{, for } 1 \le j \le n - 1$.
\end{lemma}

\begin{proof}
The $j$-th element in $\mathcal{L}_1$ is $l_j = i_j - i_{j-1}$, by definition.
By supposing that $i_j = i'_n - i'_{n-j}$,
for each $1 \le j \le n - 1$, then, $i_{j-1} = i'_n - i'_{n-j+1}$.
Therefore, $$l_j = i_j - i_{j-1} = i'_n - i'_{n-j} - i'_n + i'_{n-j+1} = i'_{n-j+1} - i'_{n-j}$$ which is equal to the $(n - j + 1)$-th element of $\mathcal{L}_2$, by definition.
\end{proof}

Furthermore, we denote by $\mathcal{L}_1 \oplus \mathcal{L}_2$ the %
increasing
sequence obtained by merging the interval-sequences of $\mathcal{L}_1$ and $\mathcal{L}_2$, and discarding repetitions.
It is not difficult to use Lemma~\ref{prop:rev-comb1} and prove that $\mathcal{L}_1 \oplus \mathcal{L}_2$ is equal to the reverse of the sequence induced by merging the interval-sequences of the reverse of $\mathcal{L}_1$ and $\mathcal{L}_2$. Then, denoting by $\mathcal{L}_1^r$ and $\mathcal{L}_2^r$ the reverse of $\mathcal{L}_1$ and $\mathcal{L}_2$, respectively, the following Lemma can be stated.

\begin{lemma}
\label{prop:rev-comb2}
Given two sequences $\mathcal{L}_1$ and $\mathcal{L}_2$ with the same number $n$ of elements,
then $\mathcal{L}_1 \oplus \mathcal{L}_2$ is equal to the reverse of $\mathcal{L}_1^r \oplus \mathcal{L}_2^r$.
\end{lemma}

\noindent
Now, Theorem \ref{prop:comb} can be proved as a consequence of Lemmas~\ref{prop:rev-comb1} and \ref{prop:rev-comb2}.
By definition, the double-stranded factorization $\fact\comb(s)$ is induced by the fingerprint $\fingerprint(s) \oplus \fingerprint(\overline{s})^r$ where $\fingerprint(s)$ and $\fingerprint(\overline{s})$ are the fingerprints of $s$ and $\overline{s}$ (respectively), obtained by applying
the basic factorization algorithm $\fact$.
Similarly, $\fingerprint(\overline{s}) \oplus \fingerprint(s)^r$ is the fingerprint inducing $\fact\comb(\overline{s})$.
By Lemma~\ref{prop:rev-comb2}, we have that
$\fingerprint(s) \oplus \fingerprint(\overline{s})^r$ (fingerprint of $s$) is equal to the reverse of $\fingerprint(s)^r \oplus (\fingerprint(\overline{s})^r)^r$.
Since $(\fingerprint(\overline{s})^r)^r$ is $\fingerprint(\overline{s})$, then $\fingerprint(s) \oplus \fingerprint(\overline{s})^r$ is the reverse of $\fingerprint(s)^r \oplus \fingerprint(\overline{s})$ (that is, the fingerprint of $\overline{s}$) and Theorem~\ref{prop:comb} is proved.

\paragraph{Computational complexity of Lyndon-based factorization algorithms.}
The computational complexity of the algorithms to compute the Lyndon-based factorizations plays a crucial role in the realization of \texttt{lyn2vec}.
As explained above, $\CFL$ and $\ICFL$ can be computed in linear time and constant space.
Regarding the computation of the $\CFLICFL$ factorization of a string $s$, our idea is to use a main process for computing $\CFL(s)$, and as soon as a $\CFL$ factor is computed, a new (parallel) process computes its $\ICFL$ factorization. In conclusion, a $\CFLICFL$ factorization can also be computed in linear time.

\paragraph{Factor length distribution in Lyndon-based factorizations and conservation property.}

The Conservation Property allows to capture the similarity between two strings. However, in the extreme case of factorizations consisting of just one factor, the Conservation Property does not apply.
Moreover, the property has been proved for $\CFL$  and its extensions to $\CFLICFL$ and $\CFLICFL\comb$ is a trivial consequence of the fact that $\CFL$ factors shared by two strings are factorized in the same way by the $\ICFL$ algorithm. Currently, we cannot prove that the Conservation Property also holds for the $\ICFL$ algorithm, nevertheless, the high accuracy of the experimental results suggests the validity of this conjecture.

\begin{figure}
	\centerline{\includegraphics[width=\textwidth]{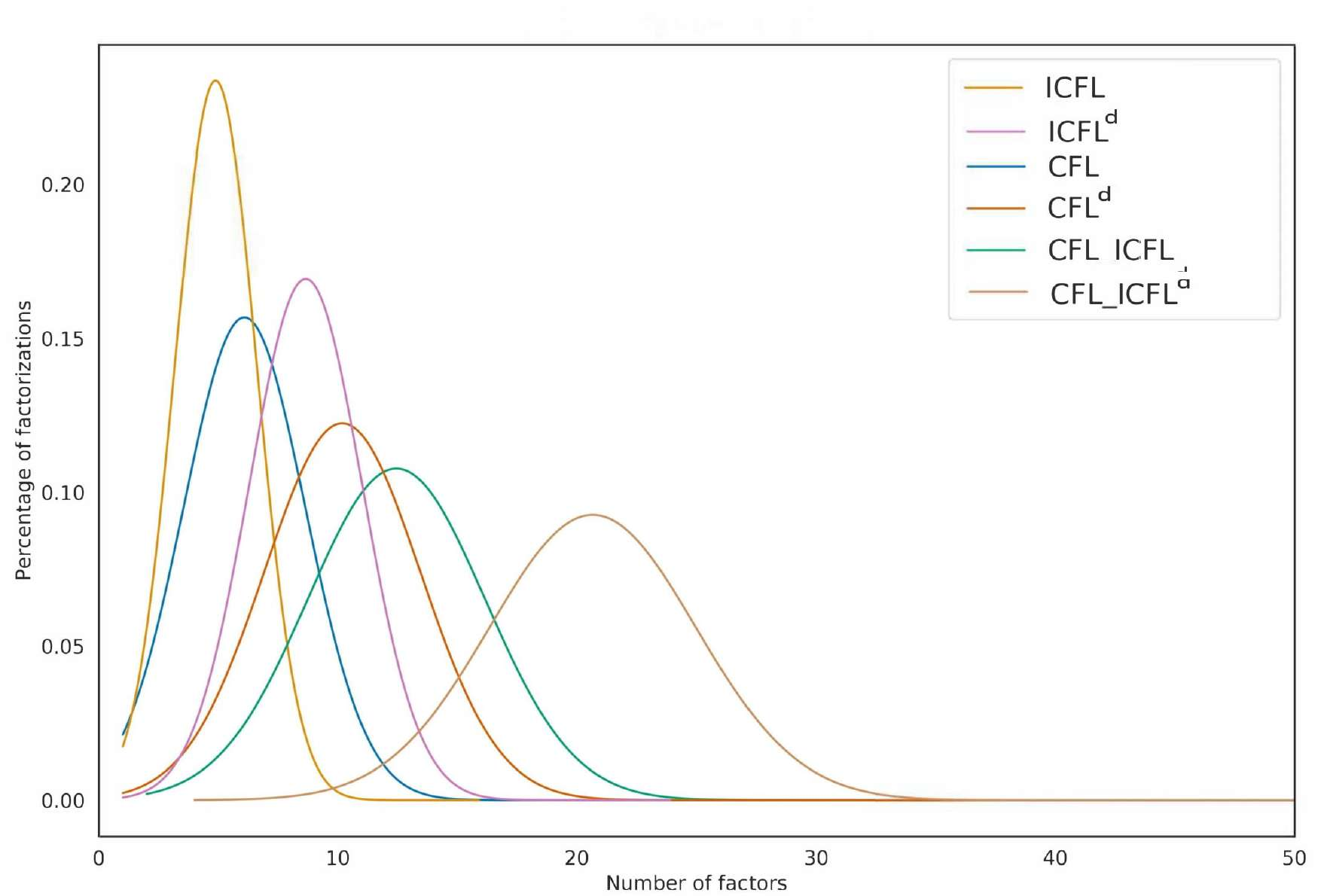}}
	\caption{Distribution of the number of factors. The factorization algorithms are listed in the legend by decreasing peak.}
	\label{fig:number-of-factors}
\end{figure}

An experimental analysis was also performed in order to investigate the distribution of the number of factors of Lyndon-based factorizations computed on a simulated error-free dataset.
Undoubtedly, a high number of factors in the factorizations fosters the possibility of detecting common regions between strings.
The dataset was simulated by using \texttt{dwgsim}~\cite{homerdwgsim} and includes 21 million 150-long genomic reads extracted from the region \texttt{960,000}-\texttt{80,960,000} of the human Chromosome 1, for a total of 80 million bases.
Figure~\ref{fig:number-of-factors} shows, for each considered factorization algorithm, the distribution of the number of factors per read.
Note that $\CFLICFL\comb$ produces about twice as much factors than any other algorithm, whereas $\CFL$, $\ICFL$, and $\ICFL\comb$ produce a small number of factors with a low distribution variance.

It is worth observing that the question on how many factors  are present in a Lyndon factorization  has been recently faced in~\cite{stacs2021}.

\subsection{Uniqueness of the fingerprint and $k$-fingers: the collision phenomenon}
\label{ssec:collisions}

We recall that our main goal is to use $k$-fingers for capturing the similarity of two sequences but unfortunately, completely distinct strings may share the same $k$-fingers or even the same fingerprint as shown in the following example.

\begin{example}
    Let $x = CCGGTT$ and $y = AACCGG$ be two strings whose $\ICFL$ factorizations are $\ICFL(x) = \langle CC,GG,TT \rangle$ and $\ICFL(y) = \langle AA,CC,GG \rangle$, respectively. Hence, $\fingerprint(x)=\langle 2,2,2 \rangle$ and $\fingerprint(y)=\langle 2,2,2 \rangle$.
\end{example}

At this point, we can introduce the definition of \emph{collision} of a $k$-finger.

\begin{definition} [$k$-finger collision]
\label{def:collision}
Let $x,y \in \Sigma^*$ be two strings and let $\fact$ be a
Lyndon-based factorization algorithm.
Let $\fingerprint(x)$ and $\fingerprint(y)$
be the fingerprints for $x$ and $y$ with respect to $F$.
Let $\mathcal{K}_x$ be a $k$-finger of $\fingerprint(x)$
and $\mathcal{K}_y$ be a $k$-finger of $\fingerprint(y)$.
Let $s_{\mathcal{K}_x}$ and $s_{\mathcal{K}_y}$ be  the two substrings of $x$ and $y$ supporting  $\mathcal{K}_x$ and $\mathcal{K}_y$, respectively.
If $\mathcal{K}_x = \mathcal{K}_y$
and $s_{\mathcal{K}_x} \neq s_{\mathcal{K}_y}$,
then we say that there exists a \emph{collision} for the $k$-finger $\mathcal{K}_x = \mathcal{K}_y$.
\end{definition}

Observe that, since a fingerprint can be viewed as a string over the alphabet of the positive integers, $k$-fingers can be considered as $k$-mers of fingerprints and collisions can be studied by exploiting the results already obtained in the literature when assigning numbers to canonical $k$-mers~\cite{mash}.

\paragraph{Collision measure.}
At this point, given a collection $S$
of strings (or reads), a Lyndon-based factorization algorithm and a value of $k$, the question is: ``How can we measure $k$-finger collisions in the obtained fingerprints?''.
A na\"ive approach could be the following. For each $k$-finger $\mathcal{K}$,
the set $S_{\mathcal{K}}$ of its supporting strings in the collection $S$ is considered.
Let us define $count(s_{\mathcal{K}})$ the number of occurrences of the string $s_{\mathcal{K}} \in S_{\mathcal{K}}$ as supporting sequence of $\mathcal{K}$.
The effective number of possible collisions
with respect to $\mathcal{K}$ (collision level
related to $\mathcal{S}$)
is given by the number of
all the different pairs of
such supporting strings, i.e.,
$\displaystyle\sum_{w_1,w_2 \in S_{\mathcal{K}}} count(w_1) \times count(w_2)$.
Clearly, the overall collision level of the fingerprints of $S$ can be simply computed as the sum of the collision levels of all the obtained $k$-fingers.

An example of collision level for a $k$-finger is illustrated in the following.

\begin{example}
Let $S$ be a set composed of just one string $s=abbbcaddabcaabddabcbbbcadd$ $abcaa$.
Let $\fact(s)=\langle a, bb, bca, dd, abc, $ $aa, b, dd, abc, bb, bca, dd, abc, aa \rangle$ be the factorization of $s$;
hence, $\fingerprint(s)=\langle 1, 2, 3, 2, 3, 2, 1, 2, 3, 2, 3, 2, 3, 2 \rangle$. The $3$-finger $\mathcal{K}=\langle 2, 3, 2 \rangle$ occurs $5$ times in the fingerprint and the set of its supporting strings is $S_{\mathcal{K}}= \{bbbcadd, ddabcaa,$ $ddabcbb\}$.
In particular,
$\mathit{count}(bbbcadd)=2$, $\mathit{count}(ddabcaa)=2$ and $\mathit{count}(ddabcbb)=1$.
So, the collision level of $\mathcal{K}$ is
$$ c_{\mathcal{K}} = \displaystyle\sum_{w_1,w_2 \in S_{\mathcal{K}}} count(w_1) \times count(w_2) =
2 \times 2 + 2 \times 1 + 2 \times 1 = 8.
$$
\end{example}

Unfortunately, such measure has some drawback when used on large collections of strings (or reads), since it may be computationally expensive and may not give any statistical information to be usefully exploited.

Therefore, we propose %
a novel metric to measure the collision level for a set of $k$-fingers, named \textit{Collision Rate}.

\begin{definition}[Collision Rate]
Let $S$ and $F$ be a set of strings and a factorization algorithm.
Given the set $L_k$ of $k$-fingers extracted from the fingerprints of $S$ and the set $S_k$ of their supporting strings, then, the \emph{Collision Rate} is $c$ $=$ $\frac{|S_k|}{|L_k|}$.
\end{definition}

\begin{example}
Let $L_k=\{\langle 2,2,2 \rangle\}$ be the set of distinct $k$-fingers extracted, and let $S_k = \{AACCGG, AACCTT, CCGGTT\}$ be the set of distinct subsequences corresponding to $k$-fingers in $L_k$. Then, the collision rate is
$c = \frac{|S_k|}{|L_k|} = \frac{3}{1} = 3$.
\end{example}

Note that the Collision Rate is a simple overall metric representing the average number of (distinct) strings supporting $k$-fingers.
When $|L_k| = |S_k|$, then each $k$-finger has a unique supporting string (that is, there is absence of collisions) and $c=1$.
A high value of $c$ means a high collision level.

In the next section, we
list some theoretical properties explaining the reasons for collisions, including how the lexicographic ordering of the characters of the alphabet may intervene.
It is worthy of note that the problem of finding the particular order of the alphabet symbols for obtaining an optimal Lyndon factorization (\textit{e.g.}, such that the number of factors is at most, or at least, a given threshold) is in general hard~\cite{stacs2021}.

\paragraph{Collision-free perspectives: the superfingerprint.}

Given a string $s$, $\al(s)$ will denote the set of symbols which are present in $s$, also referred as \emph{internal alphabet}.
Let $\Sigma$ be a finite alphabet and $(\Sigma, <) = (a_1, \ldots, a_n)$ be the alphabet $\Sigma$ which has been totally ordered, such that $a_i < a_{i+1}$ for $i = 1, \ldots, n-1$.
We denote by $\Pi_{\Sigma}$ the set of all the totally ordered alphabets which can be obtained
by considering all the possible orderings of the symbols in $\Sigma$.
For each $p \in \Pi_{\Sigma}$,
we denote by $<_p$ the lexicographic order induced by $p$.
In the following, $\fact_p$ will be used to refer to the (Lyndon-based) factorization algorithm $\fact$ relying on the particular ordering $<_p$ and $\overline{p}$ denotes the inverse of $p$ (\textit{e.g.}, $\overline{p}=(T,G,C,A)$) for $p=(A,C,G,T)$).

\begin{example}
Given $p = (A,C,G,T)$ and $p' = (C,A,G,T)$, then, $ACG <_p CAG$ and $CAG <_{p'} ACG$.
\end{example}

\begin{example}
Given $p = (A,C,G,T)$, $p' = (C,A,G,T)$ and $x=CAACAC$, then, $\ICFL_{p}(x) = \langle CAA,CAC \rangle$ and $\ICFL_{p'}(x) = \langle C,AACAC \rangle$.
\end{example}

Note that the considered ordering of the alphabet symbols may have a significant impact on the Lyndon-based factorization. Under some conditions, it is enough to alter the relative order of just two symbols in the alphabet to induce a relevant change in the factorization (see Proposition~\ref{P1}).
Indeed, two words, having the same length greater than $1$ and distinct \emph{internal alphabets}, will have distinct Lyndon-based factorizations under different orderings of the symbols of the alphabet.

\begin{proposition} \label{P1}
Let $\Sigma$ be a finite alphabet. Let $x,y \in \Sigma^*$ be such that $\al(x) \not = \al(y)$ and $|x| = |y| > 1$. For any $p \in \Pi_{\Sigma}$ there will exist $p' \in \Pi_{\Sigma}$ such that only one of the orderings of $\al(x)$ and of $\al(y)$ is changed in $p'$.
Moreover, if
$x$ and $y$ are both Lyndon (resp. inverse Lyndon)
words with respect to $ <_p$, then
only one of $x$ and $y$ is a Lyndon (resp. inverse Lyndon)
word with respect
to $ <_{p'}$.
\end{proposition}
\begin{proof}%
Let $x,y \in \Sigma^*$ be such that
$\al(x) \not = \al(y)$ and $|x| = |y| > 1$.
Set $$x = a_1 \cdots a_n, \quad y = a'_1 \cdots a'_n$$
Since $\al(x) \not = \al(y)$, one of the following two cases can occur (the other cases are symmetric)

\begin{itemize}
\item[(1)]
$a'_1 \not \in \al(x)$.
\item[(2)]
$a'_1 \in \al(x)$ and
there exists $i$, $1 < i \leq n$, such that $a'_i \not \in \al(x)$.
\end{itemize}

Assume that case (1) holds.
For any $p \in \Pi_{\Sigma}$, let
$p' \in \Pi_{\Sigma}$ be such that $a <_{p'} a'_1$, for any $a \in \Sigma \setminus \{a'_1 \}$,
whereas the order on the other symbols remains unchanged.
In particular, the ordering of $\al(x)$ is not
changed in $p'$.
Assume that
$x$ and $y$ are both Lyndon words with respect to $ <_p$.
Since $y$ is a Lyndon word and $|y| > 1$,
there exits $a'_j \in \al(y)$ with $a'_j \not = a'_1$.
Thus, $y = a'_1 \cdots a'_n >_{p'} a'_j \cdots a'_n$ and $y$ is not a Lyndon word
with respect to $ <_{p'}$.
On the contrary, since the ordering of $\al(x)$ is not changed in $p'$,
$x$ is still a Lyndon word with respect to $<_{p'}$.

Assume that case (2) holds.
For
any $p \in \Pi_{\Sigma}$, let
$p' \in \Pi_{\Sigma}$ be such that $a'_i < a$, for any $a \in \Sigma \setminus \{a'_i \}$,
whereas the order on the other symbols remains unchanged.
In particular, the ordering of $\al(x)$ is not
changed in $p'$.
Assume that
$x$ and $y$ are both Lyndon words with respect to $<_p$. Thus, $y = a'_1 \cdots a'_n >_{p'} a'_i \cdots a'_n$ and $y$ is not a Lyndon word
with respect to $ <_{p'}$.
On the contrary, the ordering of $\al(x)$ is not changed in $p'$, hence $x$ is still a Lyndon word with respect to $ <_{p'}$.
\end{proof}

\begin{example}
Words $x = ACG$ and $y = ACT$ are Lyndon words with respect to $p = (A,C,G,T)$. Note that $x$ is also a Lyndon word with respect to $p' = (T,A,C,G)$, since $<_{p'}$ maintains the same reciprocal order of the symbols of its \emph{internal alphabet} $\{A,C,G\}$.
On the contrary, word $y = ACT$ is not a Lyndon word with respect to $<_{p'}$, since the reciprocal order of the symbols of its \emph{internal alphabet} $\{A,C,T\}$ changes from $<_p$ to $<_{p'}$ and $y \succ T$.
\end{example}

The choice of a specific ordering of the alphabet symbols can have a significant impact on the collision level of the set of extracted $k$-fingers.
However, understanding whether there exists a specific order minimizing the collision level, as well as how to compute it, remains an open problem, and, as explained in Section~\ref{sec:conclusion}, it is worthy of further investigation.

At this point, we introduce the notion of \emph{superfingerprint} extending the notion of fingerprint. The superfingerprint combines the "regular" ordering $p=(A,C,G,T)$ with the inverse ordering $\overline{p}=(T,G,C,A)$ and is defined as the concatenation of the fingerprint obtained with respect to $p$ with the fingerprint obtained with respect to $\overline{p}$.

\begin{definition}[Superfingerprint]
\label{def:superfingerprint}
Given the ordered alphabets $p=(A,C,G,T)$ and $\overline{p} = (T,G,C,A)$, a string $s$ and a factorization algorithm $\fact$,
then, the superfingerprint of $s$ is the concatenation $\langle \fingerprint_{p}(s)$, $\$$,
$\fingerprint_{\overline{p}}(s) \rangle$, where $\fingerprint_{p}(s)$ is the fingerprint obtained from $\fact_p(s)$,
$\fingerprint_{\overline{p}}(s)$
is the fingerprint obtained from $\fact_{\overline{p}}(s)$ and the symbol \$ is used to mark the separation between the two fingerprints.
\end{definition}

We carried out some experiments in order to test the impact of the combined use of different orderings.
As we will see in Section~\ref{sec:evaluation}, the results suggest that, from a practical point of view, this approach reduces the impact of collisions.

\subsection{\texttt{lyn2vec}: a tool for Lyndon-based sequence embedding}
\label{ssec:lyn2vectool}

We have implemented \texttt{lyn2vec}, a novel tool for providing sequencing read embeddings relying on \textit{Lyndon-based factorizations}.
\texttt{lyn2vec} takes as input a file of sequencing reads in \texttt{FASTA} or \texttt{FASTQ} format and executes a factorization algorithm (specified as input) over each input read in order to compute its representation. Specifically, \texttt{lyn2vec} produces the following types of representations:

\begin{itemize}
\item the fingerprint
\item the sequence of the $k$-fingers extracted from the fingerprint of of the read (given a value of $k$ specified as input parameter)
\item the sequence of the $k$-fingers extracted from the superfingerprint (Definition~\ref{def:superfingerprint}), after discarding the $k$-fingers containing the separator \$.
\end{itemize}

As an example, let us consider $s = GCATCACCGCTCTACAG$.
By using the $\CFLICFL$ algorithm with a threshold $T = 30$, the factorization and the fingerprint of $s$ (with respect to the "regular" ordering $(A,C,G,T)$) will be $\CFLICFL(s)$ $=$ $\langle$ $G$, $C$, $ATC$, $ACCGCTCT$, $ACAG$ $\rangle$ and $\mathcal{L}(s) = \langle 1,1,3,8,4 \rangle$, respectively.

Observe that $\langle G, CA, TCACCGC, TCTACAG \rangle$ is the factorization with respect to the inverse ordering $(T,G,C,A)$ and the related fingerprint is $\langle 1,2,7,7 \rangle$. Then,
the superfingerprint will be given by $\mathcal{S}(s) = \langle 1,1,3,8,4,\$,1,2,7,7 \rangle$, that is, by the concatenation of the two fingerprints by interposing the separator \$.
By assuming $k=3$, \texttt{lyn2vec} will output one of the following three representations:

\begin{itemize}

\item the \textit{fingerprint} $\mathcal{L}(s) = \langle 1,1,3,8,4 \rangle$ obtained from $\CFLICFL(s)$

\item the sequence $\langle \langle 1,1,3 \rangle, \langle1,3,8 \rangle, \langle3,8,4 \rangle \rangle$ of the \textit{$k$-fingers} extracted from $\mathcal{L}(s)$

\item the sequence $\langle \langle1,1,3 \rangle$, $\langle1,3,8\rangle, \langle3,8,4\rangle, \langle1,2,7\rangle, \langle2,7,7\rangle \rangle$ of the \textit{$k$-fingers} extracted from $\mathcal{S}(s)$
\end{itemize}

\texttt{lyn2vec} has been implemented in Python by using the \textit{Scikit-learn} library\footnote{\url{https://scikit-learn.org/stable/}}. The source code and all the files produced in the evaluation tests are available online\footnote{\url{https://github.com/rzaccagnino/lyn2vec}}.

\section{Experiments} \label{sec:evaluation}

In this section, we provide the details of the experiments carried out to assess the effectiveness of \texttt{lyn2vec}, \ie its capability to provide numeric representations of biological sequences that can be effectively learned by machine learning models.

The experiments are organized in two parts. The first part (Section~\ref{sec:rgc}) focuses on the \emph{read-gene classification} problem~\cite{shark_bioinformatics} in order to perform a series of benchmark experiments.
The second part (Section~\ref{sec:gene-fusion}) is devoted to extend the \emph{rule-based} classifier, designed in the first part (for the \emph{read-gene classification} problem), in order to tackle a more complex problem, namely the detection of chimeric reads in a sample of transcriptomic (RNA-Seq) reads, which is the preliminary step for finding gene fusions~\cite{Mertens2015, Kumar2016}.

The read-gene classification problem asks for the classification of each read of a given set of RNA-Seq reads to its putative origin gene. Instead, the chimeric reads detection problem can be seen as a generalization of read-gene classification since it requires to assign a chimeric read to two fused genes (instead of a single gene).

From a computational point of view, an RNA-Seq read is a substring of a transcript (messenger RNA or mRNA) expressed from a genomic region (\emph{locus}) called \emph{a gene}.
Hence, the origin gene of a RNA-Seq read is the gene expressing the transcript which the read has been sequenced from.
We remark that read-gene classification can be efficiently solved by using aligners~\cite{minimap2,star} or pseudo-aligners~\cite{salmon2017,kallisto2016}.
However, our goal is to test, in a machine learning (ML) context, the capability of the embedding representations produced by \texttt{lyn2vec} to highlight the common regions between reads.
Observe that any ML model performing a read-gene classification would not be able to produce accurate results if the representation (or embedding) of the sequences were not able to adequately represent their content.
On the contrary, if the ML model is able to perform a sufficiently good classification, we can conclude that the embedding adequately represents the sequence content.
In the first experimental part, we investigate the feasibility of our proposed representations in a simplified scenario. We want to avoid a more complex application where aspects, such as the fine tuning of the ML model, the completeness of the data and the comprehension of the underlying biological process are crucial to achieve the utmost accuracy.

The second experimental part is focused on detecting chimeric reads in a set of RNA-Seq reads.
A chimeric read is a read sequenced from a transcript expressed by a fusion gene, that is, a hybrid gene composed of two (or more) genes joined together by a structural variation event; fusion genes are involved in many types of human neoplasia~\cite{Mertens2015}.
We show how the \emph{rule-based} classifier (Section~\ref{ssec:EQ2-task}), designed in the first part for the read-gene classification, has been extended in order to detect chimeric reads in a RNA-Seq set.

In the following sections, we assume that the reader is familiar with the basic notions of Machine Learning. However, we refer to~\cite{tan2016introduction} for further details.
\subsection{Read-gene classification}
\label{sec:rgc}

We remark that, as described in Section \ref{ssec:lyn2vectool}, \texttt{lyn2vec} provides three types of read representation. As a consequence, we faced the read-gene classification problem in three different frameworks, each one considering a representation provided by \texttt{lyn2vec}.
We organized our experiments in three groups, we call \emph{tasks}, answering the following three questions:
\begin{itemize}
    \item[\textbf{T1.}] \textit{``How effective is the fingerprint produced by \textup{\texttt{lyn2vec}} as read representation in the read-gene classification problem?''}
    \item[\textbf{T2.}] \textit{``How effective is the sequence of the $k$-fingers extracted from the fingerprints produced by \textup{\texttt{lyn2vec}} as read representation in the read-gene classification problem?''}
    \item[\textbf{T3.}] \textit{``How effective is the sequence of the} $k$-fingers extracted from the superfingerprint produced by \textup{\texttt{lyn2vec}} as read representation in the read-gene classification problem?''

\end{itemize}

Each task is composed of sub-tasks (or \emph{experiments}), each one devoted to (find and) assess a classifier using a specific representation.

In task T1, the fingerprints of the reads are directly used as \textit{feature vectors} (each fingerprint will correspond to one feature vector) on which the Machine Learning model will be trained.
We remark that, in a classification problem, the feature vector represents the list of features of an object considered relevant for training a model to classify the object itself.
In tasks T2 and T3, the $k$-fingers (extracted from the fingerprints or superfingerprints) are used as \emph{feature vectors}. Then, (i) the
model is first trained in order to assign each $k$-finger to a gene and next, (ii) a special classifier (Section~\ref{ssec:EQ2-task})
exploits the obtained result in order to assign each read to a class (or gene).

As shown below, the fingerprint representation can work well in case of error-free data, while we observe a decrease in the performance when sequencing errors are present. On the other hand, the $k$-finger-based representation turns out more robust when errors are present.

Finally, we have empirically proved that the $k$-finger-based representation allows to reduce the collisions and improve performances (Section~\ref{ssec:EQ3-task}). However, we remark that the problem of choosing the best ordering of the alphabet to reduce collisions is still open.

\paragraph{Data Setting.}
We used the annotation (\texttt{havana} and \texttt{ensembl\_havana}) of the $6040$ human genes of chromosomes $1$, $17$ and $21$, containing a total amount of $17,314$ transcripts; we randomly selected $100$ genes out of them in order to obtain a small set of genes for assessing the effectiveness of our embedding representation. For each one of the 100 genes, 4  transcripts were randomly selected (for a total number of 400 transcripts). Then, from each one of such 400 transcripts, we extracted all the $100$-mers, thus obtaining a total of $797,407$ $100$-long substrings. We used these substrings as \emph{error-free reads}
from which to extract the feature vectors used in the following experiments and
to carry out the analysis presented in Section~\ref{sec:error-analysis}. Each \emph{read} clearly belongs to one of the considered $100$ classes (one class for each gene).

\paragraph{Feature Extraction.} The datasets of feature vectors are extracted from the collection of 797,407 100-long \textit{input reads} of our data set. We recall that each fingerprint corresponds to a feature vector in task T1 and each $k$-finger extracted from a fingerprint (resp. superfingerprint) corresponds to a feature vector in task T2 (resp. T3). Fingerprints and $k$-fingers will be also referred as \emph{samples}.
In more detail, a total of $10$ factorization algorithms were considered: the four algorithms $\CFL$, $\ICFL$, $\CFL\comb$ and $\ICFL\comb$ plus the two algorithms $\CFLICFL$ and $\CFLICFL\comb$ applied for the three values $\{10, 20, 30\}$ of the parameter $T$.
For each algorithm, we have computed (by using \texttt{lyn2vec}) the fingerprint and the superfingerprint  of each input read, thus obtaining $10$ datasets of fingerprints ($10$ experiments of task T1) and $10$ datasets of superfingerprints ($10$ experiments of task T3 where the feature vectors are the $k$-fingers extracted from the superfingerprints).
Next, the \emph{$k$-fingers} for $k$ from $3$ to $8$ (a total of six values) are extracted from the $10$ datasets of fingerprints, thus obtaining a total of $60$ datasets of $k$-fingers ($60$ experiments of task T2) and $60$ datasets of $k$-fingers extracted from superfingerprints ($60$ experiments of task T3).
Since the length of a feature vector must be constant and the number of elements (integer values) composing our feature vectors (fingerprints or $k$-fingers) is clearly variable, then, we need to perform a padding with trailing $-1$s in order to achieve a constant size over all the feature vectors. Observe that a $k$-finger may even be padded, when the read fingerprint is shorter than $k$.

\paragraph{Basic Methodology.} For each one of our experiments, the following steps are performed in order to produce a classifier:
\begin{enumerate}
	\item
	\textit{Labeling.} Each feature vector (fingerprint or $k$-finger), in the dataset, is labeled in order to create a link with its class (that is, with its origin gene).
	\item
	\textit{Learning.}
	The following ML models are considered: \textit{Random Forest} (\texttt{RF}), \textit{Logistic Regression} (\texttt{LR}), and \textit{Multinomial Naive Bayes} (\texttt{MNB}). Each model is trained by using the labeled samples.

Next, the \textit{k-fold cross-validation} technique (\ie finding the optimal hyperparameter values yielding a satisfying generalization performance) is used in combination with the \textit{GridSeachCV} method in order to select the best model and obtain a classifier.
The dataset is first split in two subsets using \textit{stratification}\footnote{The stratification has been performed by using the method \texttt{train\_test\_split} of \textit{Scikit-learn} Python library.}: a \textit{training set} (80\% of the samples) and a \textit{testing set} (20\% of the samples). \emph{Stratification} guarantees that all the considered $100$ classes (genes) are represented in both sets, while maintaining the same proportions of the original dataset. Moreover, to minimize data leakage, we have used a common ML practice consisting in applying normalization within the cross-validation folds, separately. Specifically, we have normalized the data by using the \textit{MinMaxScaler} technique. The normalization is used to map the values of the features (integer values in our case) to a fixed range (usually $[0,1]$), providing better results with respect to the case of features having values in different ranges.
We remark that \emph{stratification} and \emph{normalization} are two well-known ML techniques.
Next, the $k$-fold cross validation is performed on the training set (we tried several values of $k$ and $5$ gives good results), and, at each one of the $k$ steps (at each step, $k-1$ folds are used for training the models, and one fold, named \textit{test fold}, is used for performance evaluation), we apply an exhaustive search with specified parameters (GridSeachCV method), and for each combination of such values, we compute the average performance of the considered classifier reached on the current independent test fold. Finally, after discovering satisfactory hyperparameter values, we perform a retraining of the best model (the model showing the best performance) on the training set, thus producing a classifier.
\item
\textit{Evaluation.}
The \textit{generalization capability} of the obtained classifier, \ie the capability to reach high performance on unseen samples, is assessed by applying it for classifying the elements of the \textit{Testing Set} using the best parameters found in the previous step.
\end{enumerate}

\paragraph{Testing on simulated RNA-Seq reads.}
We also performed a test over a simulated RNA-Seq dataset. We simulated a total of $10$ million $100$-long reads from the $17,314$ transcripts of our data setting by running Flux Simulator~\cite{fluxsimulator} for different expression levels. Default parameters and the default Illumina error model were used; the generation of Poly-A tails was disabled.
Then, only the $285,628$ reads related to our panel of $100$ genes were retained and used for evaluating the effectiveness of \texttt{lyn2vec} in classifying RNA-Seq reads from fingerprints (Section~\ref{ssec:EQ1-task} and from $k$-fingers (Section~\ref{ssec:EQ2-task})).
Observe that this set is unbalanced: $142,266$ reads were simulated from gene \texttt{ENSG00000132517} (the most expressed in the dataset), whereas only $2$ reads were simulated from gene \texttt{ENSG00000116205} (the least expressed in the dataset).
We point out that, since a fingerprint represents an entire read, then the class of a read is trivially the class of its fingerprint and a classifier of fingerprints is a classifier of reads.
On the contrary, when the reads are represented by their $k$-fingers, then a classifier of $k$-fingers must be accompanied by a \emph{rule-based} read classifier (Section~\ref{ssec:EQ2-task}) for inferring the class of a read from the classes of its $k$-fingers.

\paragraph{Performance scores.}
For each experiment, the results obtained in the \emph{Evaluation} step are evaluated by using a set of metrics (usually) defined for binary classification problems (that is, only two classes are taken into account). Precisely, \emph{accuracy} = $\frac{TP+TN}{TP+TN+FP+FN}$, \emph{precision} = $\frac{TP}{TP+FP}$, \emph{recall} = $\frac{TP}{TP+FN}$ and \emph{F-score} = $2 \cdot \frac{precision \cdot recall}{precision + recall}$. In short, the accuracy measures the portion of correctly classified samples, the precision indicates how many samples assigned to a class actually belong to such a class, and the recall gives the amount of samples of a given class which are correctly assigned to that class. Finally, the F-score is a useful summary metric defined as the harmonic mean of precision and recall.
These metrics are also applied to evaluate the classification results (Sections~\ref{ssec:EQ1-task} and~\ref{ssec:EQ2-task}) obtained from the classification of the set of the simulated RNA-Seq reads.

In a multi-class framework, the global performance of a classifier can be obtained by first averaging the metrics computed for each single class, and then considering the arithmetic or the weighted mean. In this work, since we also performed a testing over a dataset of RNA-Seq reads, which is unbalanced, we used the averaged values of these metrics weighted for the number of actual samples belonging to each class.
Sections~\ref{ssec:EQ1-task}, \ref{ssec:EQ2-task} and~\ref{ssec:EQ3-task} report the performance results (both for the \emph{Evaluation} step and the testing over the simulated dataset of RNA-Seq reads) by grouping the experiments by task. Section~\ref{sec:error-analysis} presents the results of a study about the behavior of the classification errors obtained in the best-performing experiment.

\subsubsection{Effectiveness of the fingerprint representation (task T1)}
\label{ssec:EQ1-task}

Task T1 is composed of $10$ experiments, each one related to a dataset of fingerprints computed with a specific factorization algorithm (see paragraph \emph{Feature Extraction} of Section~\ref{sec:rgc}); the read fingerprints are the samples training the considered ML models (\emph{Training} step).
As a result, we have observed that the \texttt{RF} model always outperforms the other models and we only report its results on the testing set (see Table~\ref{tab:EQ1-classification}).
As the table shows, algorithm $\CFLICFL\comb$ gives in general the best results, and the best classifier is for the parameter $\texttt{T} = 20$.

\begin{table}[h!t]
\begin{center}
\fontsize{2.5mm}{2.5mm}\selectfont{
		\renewcommand{\arraystretch}{1.1}
		\setlength\tabcolsep{2pt}
\begin{tabular}{ccllll}
\toprule
\multicolumn{2}{l}{\textbf{Experiment}}                 & \textbf{Accuracy} & \textbf{Precision} & \textbf{Recall} & \textbf{F-score} \\
\midrule
\multicolumn{2}{l}{\multirow{1}{*}{$\CFL$}}                               & 0.43             & 0.45                 & 0.43              &      0.43         \\
\midrule
\multicolumn{2}{l}{\multirow{1}{*}{$\ICFL$}}                    & 0.40             & 0.45                & 0.40              & 0.41
\\

\midrule
\multirow{3}{*}{$\CFLICFL$}        & \multirow{1}{*}{\texttt{T=10}}               & 0.90             & 0.90                 & 0.90              & 0.90               \\

                                   & \multirow{1}{*}{\texttt{T=20}}                & 0.90             & 0.90                 & 0.90              & 0.90               \\

                                   & \multirow{1}{*}{\texttt{T=30}}               & 0.89             & 0.89                 & 0.89              & 0.88               \\

\midrule
\multicolumn{2}{l}{\multirow{1}{*}{$\CFL\comb$}}                               & 0.72             & 0.72                 & 0.72              &      0.72         \\
\midrule
\multicolumn{2}{l}{\multirow{1}{*}{$\ICFL\comb$}}                    & 0.85             & 0.85                & 0.85              & 0.85
\\

\midrule
\multirow{3}{*}{$\CFLICFL\comb$}        & \multirow{1}{*}{\texttt{T=10}}               & 0.92             & 0.93                 & 0.92              & 0.93               \\

                                   & \multirow{1}{*}{\texttt{T=20}}                & \textbf{0.93}             & \textbf{0.94}                 & \textbf{0.93}              & \textbf{0.94}               \\

                                   & \multirow{1}{*}{\texttt{T=30}}               & 0.92             & 0.93                 & 0.92              & 0.92               \\

 \bottomrule

\end{tabular}}
\end{center}
\caption{\footnotesize{Performance results of the \texttt{RF} model in the $10$ experiments of task T1 (\emph{Evaluation} step})}
\label{tab:EQ1-classification}
\end{table}

To assess the effectiveness of \texttt{lyn2vec} we compared its results with those obtained by using the embedding techniques \texttt{BioVec, fastDNA} and \texttt{DNABERT}. We compared the results obtained from such embedding representations, in the specific context of the read-gene classification problem, with the results obtained with the representations provided by \texttt{lyn2vec}. \texttt{BioVec} and \texttt{fastDNA} were both re-trained  on the training set used in task T1.
Instead, in the case of \texttt{DNABERT}~\cite{DNABERT}, the representations obtained using the proposed pre-trained models provide general comprehension of the DNA language and can be used to solve generic sequence-related tasks.
Thus, they can also be used to face the read-gene classification problem. Moreover, they stressed that the pre-training of the \texttt{DNABERT} model is resource intensive, and so the pre-trained models are presented as one of the major contributions of the paper. According to these premises, we directly used such general pre-trained models to generate the representations used in our comparison.
In particular, we have performed a further experiment on the same data set obtained with $\CFLICFL\comb$, $T=20$.
Afterwards, we once again performed the classification task by feeding the new embedded features to the RF model which was the one providing best results. Furthermore, Table~\ref{tab:comparison} shows that the performance, obtained with the other representations, is systematically lower than that obtained by using \texttt{lyn2vec}.

 \begin{table}[h!t]
\begin{center}
\fontsize{2.5mm}{2.5mm}\selectfont{
		\renewcommand{\arraystretch}{1.1}
		\setlength\tabcolsep{2pt}
\begin{tabular}{ccllll}
\toprule
\multicolumn{2}{l}{\textbf{Embedding}}                 & \textbf{Accuracy} & \textbf{Precision} & \textbf{Recall} & \textbf{F-score} \\
\midrule
\multicolumn{2}{l}{\multirow{1}{*}{\texttt{lyn2vec}}}                               & 0.93             & 0.94                 & 0.93              &      0.94         \\
\midrule
\multicolumn{2}{l}{\multirow{1}{*}{\texttt{BioVec}}}                    & 0.83             & 0.85                & 0.83              & 0.83
\\

\midrule
\multicolumn{2}{l}{\multirow{1}{*}{\texttt{fastDNA}}}                   & 0.67             & 0.65                & 0.67              & 0.67
\\

\midrule
\multicolumn{2}{l}{\multirow{1}{*}{\texttt{DNABERT}}}                   & 0.37             & 0.36               & 0.35              & 0.36
\\

\bottomrule

\end{tabular}}
\end{center}
\caption{\footnotesize{Performance results of the \texttt{RF} model with the embeddings by \texttt{lyn2vec} ($\CFLICFL\comb$ and $\texttt{T}= 20$), \texttt{BioVec}, \texttt{fastDNA} and \texttt{DNABERT}. The first row is the copy of the corresponding row of Table~\ref{tab:EQ1-classification} for $\CFLICFL\comb$ and $\texttt{T}= 20$. }}
\label{tab:comparison}
\end{table}

As depicted in Table~\ref{tab:comparison}, \texttt{lyn2vec} outperforms the other embeddings. In addition, \texttt{lyn2vec} produces a dataset of fingerprints much smaller than the ones produced by the other tools, whose embedding vectors have a fixed size.

Finally, we carried out a classification test over the set of the simulated RNA-Seq reads by using the best classifier obtained in task T1, that is, the \texttt{RF} model trained with the fingerprints from $\CFLICFL\comb$ and $T=20$, obtaining a precision of $0.85$, a recall of $0.42$ and an F-score of $0.55$. The presence of sequencing errors in the simulated reads leads to fingerprints, which differ from those used during the training of the classifier.

\subsubsection{Effectiveness of the \texorpdfstring{$k$}{k}-finger-based representation (task T2)}
\label{ssec:EQ2-task}
Task T2 involves $60$ experiments, each one related to a dataset of $k$-fingers
extracted from the $10$ datasets of fingerprints (task T1) for $k=3,4,5,6,7,8$; the $k$-fingers are (in task T2) the samples training the ML models we considered (Training step).
The goal of this task is to assess the effectiveness of $k$-fingers as feature vectors improving the classification performance when sequencing errors are present in the reads.

\begin{figure}
	\centering
    \includegraphics[width=\linewidth]{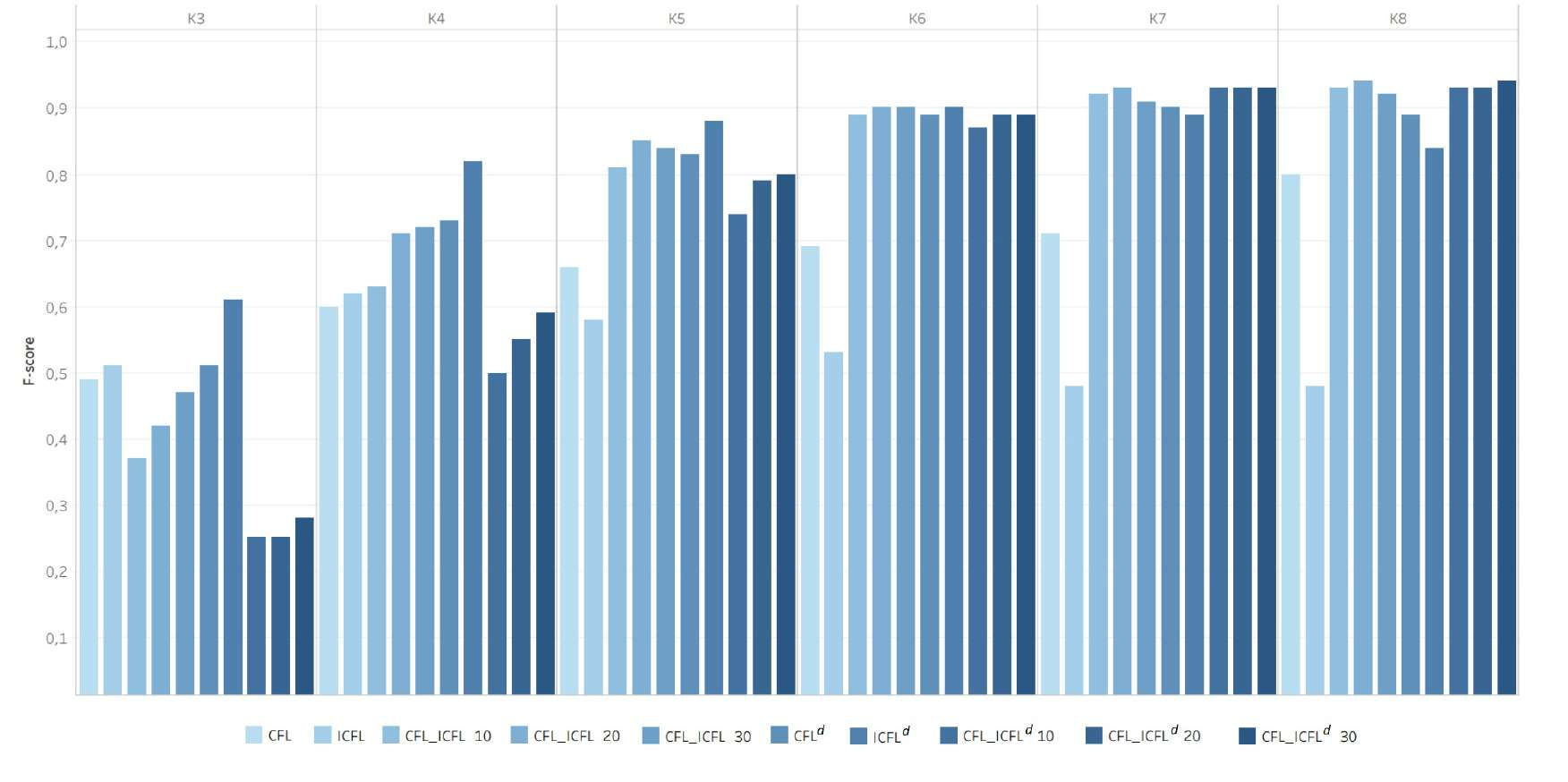}
	\caption{\footnotesize{F-score of the $60$ experiments of task T2 (\texttt{RF} model).}}
	\label{fig:EQ2-fscore}
\end{figure}

As for task T1, we only report the results for the \texttt{RF} model, since it outperforms the other models. Figure~\ref{fig:EQ2-fscore} provides an overall picture of the F-score for the total of the $60$ experiments of task T2: experiments conducted with $k=3,4,5$ show that the classifiers obtained from algorithm $\CFLICFL\comb$ (evaluated in the \emph{Evaluation} step) are the least accurate in classifying $k$-fingers.

Table~\ref{tab:EQ2-classification1} and~\ref{tab:EQ2-classification2}
report the accuracy results for the experiments with the basic factorization algorithms and with their double-stranded version, respectively. Only values of $k$ in $\{3,6,8\}$ are shown, since we observed that the classification performance tends to increase with increasing $k$. The best results (see the bold rows in the two tables) have been obtained by using $k = 8$, the factorization algorithm $\CFLICFL$ with $\texttt{T} = 20$ and its double-stranded version $\CFLICFL\comb$ with $\texttt{T} = 30$.

\begin{table}[h!t]
\begin{center}
\fontsize{2.5mm}{2.5mm}\selectfont{
		\renewcommand{\arraystretch}{1.1}
		\setlength\tabcolsep{2pt}
\begin{tabular}{cccllll}
\toprule
\multicolumn{2}{l}{\textbf{Experiment}}                 & \textbf{k} & \textbf{Accuracy} & \textbf{Precision} & \textbf{Recall} & \textbf{F-score} \\
\midrule
\multicolumn{2}{l}{\multirow{3}{*}{$\CFL$}}                   & 3                & 0.53             & 0.61                 & 0.43              &      0.49         \\
\multicolumn{2}{l}{}                                       & 6                & 0.73             & 0.75                 & 0.66             & 0.69               \\
\multicolumn{2}{l}{}                                       & 8                & 0.79             & 0.81                & 0.78              & 0.80               \\
\midrule
\multicolumn{2}{l}{\multirow{3}{*}{$\ICFL$}}                  & 3                & 0.57             & 0.60                & 0.46              & 0.51               \\
\multicolumn{2}{l}{}                                       & 6                & 0.71             & 0.77                 & 0.47              & 0.53               \\
\multicolumn{2}{l}{}                                       & 8                & 0.81            & 0.80                 & 0.42              & 0.48               \\

\midrule
\multirow{9}{*}{$\CFLICFL$}        & \multirow{3}{*}{\texttt{T=10}} & 3                & 0.47             & 0.58                 & 0.31              & 0.37               \\
                                   &                       & 6                & 0.91             & 0.91                 & 0.87              & 0.89               \\
                                   &                       & 8       &
                                   0.93    & 0.94       & 0.93     & 0.93     \\
                                   \cmidrule{2-7}
                                   & \multirow{3}{*}{\texttt{T=20}} & 3                & 0.50             & 0.57                 & 0.36     &     0.42               \\
                                   &                       & 6                & 0.92             & 0.92                 & 0.89             & 0.90               \\
                                   &                       & \textbf{8}       & \textbf{0.95}    & \textbf{0.95}        & \textbf{0.93}     & \textbf{0.94}      \\
                                   \cmidrule{2-7}
                                   & \multirow{3}{*}{\texttt{T=30}} & 3                & 0.55             & 0.62                 & 0.41              & 0.47               \\
                                   &                       & 6                & 0.91             & 0.92                 & 0.88              & 0.90               \\
                                   &                       & 8       &
                                   0.92              & 0.93                      & 0.91           & 0.92      \\
 \bottomrule

\end{tabular}}
\end{center}
\caption{Performance results of the experiments of task T2  (\emph{Evaluation} step) with the \emph{basic} factorization algorithms and $k=3,6,8$ (\texttt{RF} model).}
\label{tab:EQ2-classification1}
\end{table}

\begin{table}[h!t]
\begin{center}
\fontsize{2.5mm}{2.5mm}\selectfont{
		\renewcommand{\arraystretch}{1.1}
		\setlength\tabcolsep{2pt}
\begin{tabular}{cccllll}

\toprule
\multicolumn{2}{l}{\textbf{Experiment}}                 & \textbf{k} & \textbf{Accuracy} & \textbf{Precision} & \textbf{Recall} & \textbf{F-score} \\
\midrule

\multicolumn{2}{l}{\multirow{3}{*}{$\CFL\comb$}}            & 3                & 0.56             & 0.63                 & 0.45              & 0.51      \\
\multicolumn{2}{l}{}                                       & 6                & 0.90             & 0.91                 & 0.86              & 0.89               \\
\multicolumn{2}{l}{}                                       & 8
& 0.90          & 0.93                      & 0.86              & 0.89      \\
\midrule
\multicolumn{2}{l}{\multirow{3}{*}{$\ICFL\comb$}}  & 3                & 0.65             & 0.68                 & 0.57              & 0.61               \\
\multicolumn{2}{l}{}                                       & 6                & 0.92             & 0.92                 & 0.87              & 0.90              \\
\multicolumn{2}{l}{}                                       & 8                & 0.87             & 0.91                 & 0.78              & 0.84               \\
\midrule
\multirow{9}{*}{$\CFLICFL\comb$} & \multirow{3}{*}{\texttt{T=10}} & 3                & 0.38            & 0.55                 & 0.21              & 0.25               \\
                                   &                       & 6                & 0.89             & 0.89                & 0.86              & 0.87               \\
                                   &                       & 8                      & 0.93    & 0.93        & 0.92              & 0.93      \\
                                   \cmidrule{2-7}
                                   & \multirow{3}{*}{\texttt{T=20}} & 3                & 0.40             & 0.55                 & 0.23              & 0.25               \\
                                   &                       & 6                & 0.91             & 0.91                 & 0.88              & 0.89             \\
                                   &                       & 8                      & 0.94    & 0.94                            & 0.92     &   0.93    \\
                                   \cmidrule{2-7}
                                   & \multirow{3}{*}{\texttt{T=30}} & 3                & 0.42             & 0.51                 & 0.24              & 0.28               \\
                                   &                       & 6                & 0.91             & 0.91                 & 0.88              & 0.89             \\
                                   &                       & 8       &
                                   \textbf{0.94}                & \textbf{0.94}        & \textbf{0.93}     & \textbf{0.94} \\
\bottomrule
\end{tabular}}
\caption{Performance results of the experiments of task T2 (\emph{Evaluation} step) with the \emph{double-stranded} factorization algorithms and $k=3,6,8$ (\texttt{RF} model).}
\label{tab:EQ2-classification2}
\end{center}
\end{table}

As for task T1, we carried out a classification test over the set of the simulated RNA-Seq reads, by classifying the $k$-fingers extracted from the fingerprints of such reads. To this purpose, we have designed a \textit{rule-based classifier} in order to infer the class of a read from the classes of its $k$-fingers, that is composed of the following two cascade criteria (selected after some preliminary tests):
\begin{itemize}
    \item \textit{Majority.} Given a read, then a gene $G$ reaches the majority for that read, if at least half of its $k$-fingers are assigned to $G$; therefore, the read is assigned to $G$.
    \item \textit{Threshold.} Given a read, for each of its $k$-fingers, the \textit{classification margin} is computed; the read is assigned to the gene for which the highest value of the margin is achieved. The \textit{classification margin} is obtained by subtracting the lowest probability, whereby a $k$-finger was correctly classified during the training step, to the probability by which it was currently classified.
\end{itemize}
If \textit{Majority} (applied first) does not achieve a result, then, \textit{Threshold} will be applied.

We achieved the best classification results by using the \texttt{RF} model, trained with the $k$-fingers produced by $\ICFL\comb$ and $k = 5$; recall that the best classifier of $k$-fingers has been proved to be the \texttt{RF} model with $\CFLICFL\comb$, $\texttt{T} = 30$ and a value of $k = 8$ (high value). This behavior could be explained by the fact that a lower value of $k$ produces a higher number of $k$-fingers (for a given read), thus, a lower percentage of the extracted $k$-fingers are likely to be affected by sequencing errors.

We obtained a precision of $0.91$, a  recall of $0.77$, and a F-score of $0.82$, instead of precision of $0.85$, a  recall of $0.42$, and a F-score of $0.55$ obtained in task T1, by directly using the entire fingerprint as feature vector. Such a huge different in the performance can be explaining again by the fact that, in presence of errors, $k$-fingers are able to extract the parts of the fingerprint which are not corrupted, while in the case of the direct classification of the fingerprint, the classifier must infer the read class relying only on a (probably) single corrupted feature vector.

\subsubsection{Superfingerprints to overcome the impact of \texorpdfstring{$k$}{k}-finger collisions on the read classification (task T3)}
\label{ssec:EQ3-task}

We repeated the $60$ experiments carried out in task T2 by using the $k$-fingers extracted from superfingerprints (Definition~\ref{def:superfingerprint}) in place of the $k$-fingers extracted from fingerprints, achieving a significant improvement both in the \emph{Evaluation} step (of the $k$-finger classification) and in the testing on the simulated RNA-Seq reads.

Specifically, the \texttt{RF} model trained on $k$-fingers extracted from superfingerprints, computed with $\CFLICFL\comb$ and $\texttt{T} = 30$ and considering $k = 8$, achieves an accuracy of $0.97$, a precision of $0.96$, a recall of $0.95$, and an F-score of $0.95$ (\emph{Evaluation} step). Recall that the counterpart values of task T2 (which are the values obtained with the best-performing classifier trained with $k$-fingers from fingerprints) are $0.94$, $0.93$, $0.94$ and $0.94$, respectively.
Moreover, the test over the simulated RNA-Seq reads has been performed by exploiting the \texttt{RF} model trained on the $k$-fingers extracted from superfingerprints computed with $\ICFL\comb$ and $\texttt{T} = 30$ and considering $k = 5$; we obtained a precision of $0.95$, a recall  of $0.83$, and an F-score of $0.88$, against $0.91$, $0.77$ and $0.82$ obtained in task T2.

Since the superfingerprints cannot certainly reduce the rate of collisions, the performance improvement with respect to task T2 can only be ascribed to the fact that superfingerprints are able to reduce the impact of collisions on the classification.
This can be explained by noting that, even though additional collisions may even occur in the case when using $k$-fingers, extracted from the fingerprints built on the inverse alphabet, nevertheless the concatenation of the two fingerprints reduces the overall probability of collision, since the probability that both fingerprints are affected by a collision is lower than the probability of collision of the two separated fingerprints. In terms of how the rule-based classifier works, this means that, even if the absolute number of $k$-fingers, producing collisions, may increase, their incidence on the total number of obtained $k$-fingers will be lower, leading to a higher number of correct classifications. Still, the theoretical demonstration of the capability of superfingeprints to reduce the impact of collisions remains open. Task T3 is a first attempt for empirically assessing their effectiveness.

\subsubsection{Error analysis}
\label{sec:error-analysis}

We carried out an analysis to investigate the errors in the $k$-finger-based classification of a set of reads by using the \emph{error-free} reads of our \emph{Data Setting} (Section~\ref{sec:rgc}), thus discarding the noise due to possible sequencing errors.
Assigning a read to the wrong gene (misclassification error) is due to one of the following three reasons: \textit{(i)} the read shares substrings with transcripts coming from the wrong gene, \textit{(ii)} some $k$-fingers of the read collide (Section~\ref{ssec:collisions}) with $k$-fingers of reads coming from the wrong gene and \textit{(iii)} some $k$-fingers of the read are similar to $k$-fingers from transcripts of the wrong gene (\emph{e.g}, $\langle 2,5, 8,7,20 \rangle$ is similar to $\langle 2,5,7,7, 19 \rangle$) and the trained model produces wrong $k$-finger classifications.
We considered the results obtained in task T2 for the \texttt{RF} model trained on the $6$ datasets of $k$-fingers produced with $\CFLICFL\comb$, $T=30$ and $k$ from $3$ to $8$ (Section~\ref{ssec:EQ2-task}); in particular, recall that $k=8$ achieved the best result in task T2.
The $6$ datasets of $k$-fingers (training set plus testing set, in order to consider also the errors produced by the model in the training set) were given as input to the \emph{rule-based} classifier (Section~\ref{ssec:EQ2-task}) in order to infer the class of the (error-free) reads from which the $k$-fingers were obtained.

For each one of the $6$ classification results, we carried out the following analysis. Given the set $\mathcal{G}$ of the $100$ genes of our data setting, we considered all the pairs $(G_T, G_A)$ of the Cartesian product $\mathcal{G}^2$, such that $G_T \ne G_A$ (obtaining a total of $9900$ pairs). $G_T$ and $G_A$ will be referred as \emph{target gene} and  \emph{assignment gene}, respectively. For each $(G_T, G_A)$ we computed the number of reads originated from $G_T$ but assigned to $G_A$, referred in the following as \emph{misclassification number}.

Overall, we observed a high number of pairs having a misclassification number below $30$ (on average, $9832$ over the $6$ datasets).

The most interesting results have been obtained for $k = 5$ and $k=8$.

\paragraph{k=5} Only $29$ pairs (involving $27$ distinct genes) have a misclassification number over $30$.
The number of exchanges ranged from 0 to 158.
On average, $20$ distinct $k$-fingers per pair are misclassified. In general, about $65$\% of the misclassified $k$-fingers are shared between the two genes, and only the $15$\% of them actually correspond to common substrings, whereas the other ones were misclassifed due to reason \textit{(ii)} ($k$-finger collisions).
Not surprisingly, this result suggests that for higher values of $k$, the collision rate is lower.

\paragraph{k=8} Only $229$ pairs (involving $34$ distinct genes) have a misclassification number over $30$. The number of exchanges ranged from 0 to 113.  On average, we have found that $36$ distinct $k$-fingers per pair are misclassified due to reason \textit{(iii)} (\ie numerically close $k$-fingers).
These results suggest that \texttt{lyn2vec} provides good string representations.

This analysis highlights how different values of the parameter $k$ lead to a different behavior of the $k$-fingers and, consequently, to different results. $k=8$ seems to be the  best choice, achieving a precision and recall of 0.94 and 0.93, respectively. Limitations of the choice of higher $k$ are related to the fact that the method can misclassify similar $k$-fingers. Clearly, the length of the reads could be also another parameter to take into account to allow higher values of $k$.

These results suggest that different parameters and models applied in combination with the \texttt{lyn2vec} representation could lead to different uses whose investigation can be carried out in future works.

\subsection{Chimeric reads detection}
\label{sec:gene-fusion}

In Section~\ref{sec:rgc} we have proved the soundness of the \texttt{lyn2vec} representation in a ML-based methodology to solve the  \textit{read-gene classification} problem. This suggests the possibility of exploring \texttt{lyn2vec} in a more complex framework such as the gene fusion, where chimeric reads, originating from fused genes, are present in a set of RNA-Seq reads.
Gene fusion is the chromosome rearrangement mechanism by which two (or more) genes are joined together into a single gene (\emph{fusion gene}) and is often associated to cancer.
A \emph{chimeric} transcript originating from a fusion gene will be given by the concatenation of many parts, each one originated from one of the genes joined together. We only consider gene fusions involving two genes $g_1$ and $g_2$ and a \emph{chimeric read} will be a read $w$ sequenced from a chimeric transcript and will be composed of two parts $w_1$ and $w_2$, such that $w=w_1 w_2$ and $w_1$ and $w_2$ are originated from $g_1$ and $g_2$ respectively.

Computational tools for gene fusion detection using RNA-Seq reads have been proposed in the literature ~\cite{Wang2012, Kumar2016}.
These tools are based on a preliminary step that consists in aligning short RNA-Seq reads~\cite{Kim2019, Chiu2019, Davidson2015, Fotakis2020}
to an annotated reference genome (or transcriptome) in order to detect chimeric reads and next infer the fused genes from these reads.
Unfortunately, short reads capture relatively small pieces of the full-length gene transcripts and alignment-based approaches may not perform well, thus introducing bias in the detection of chimeric reads. On the contrary, RNA-Seq long reads are more informative to detect gene fusions, since they are composed of tens of thousands of bases and can even cover an entire transcript.
However, the exploration of methods for detecting gene fusions from a set of long reads is in its infancy.
In this regard, we conjecture that a Machine Learning approach, learning gene features, could help in this direction and we have adapted the ML-based approach, developed for the \textit{read-gene classification}, to implement an (alignment-free) method to detect chimeric reads in a set of long reads.
More precisely, given a set of candidate genes that may be involved in gene fusion events, our goal is to detect the reads which are chimeric; in other words, this problem is a rewording of the read-gene classification problem, where a chimeric read will be assigned to two genes instead of one. To this aim, we have adopted a similar strategy by first classifying the read $k$-fingers and then, inferring the two fused genes by means of a variant of the \emph{rule-based} classifier, which has been adapted to this problem.

\subsubsection{Method}

Our method for detecting chimeric reads consists of the following three steps:
\begin{enumerate}
	\item given the set of the genes of interest, the $k$-fingers are extracted from their annotated transcripts in order to train the ML model. At the same time, a set $\cal F$ of \emph{target} strings is obtained and used in the second step as a "guidance" to factorize the long reads,
	\item  for each input read, a factorization based on the set $\cal F$ is computed (referred as \emph{adaptive} factorization),
	\item from each \emph{adaptive} factorization, the $k$-fingers are extracted and classified by the trained model; next, a \emph{rule-based} classifier decides whether the read is chimeric and (in such case) determines the two involved genes.
\end{enumerate}

The following three paragraphs provide the details of the three steps.

\paragraph{Training the model.}
For each annotated transcript (of the the genes of interest), all the substrings of a given length $W$ (input parameter) are extracted and for each substring, the factorization with algorithm $\CFLICFL\comb$ with $T=30$ was computed. Next, the $k$-fingers for $k=8$ were extracted and exploited for training the ML model (by following the same methodology described for the read-gene classification problem).
This step also computes a set $\cal F$ of $L$-long substrings ($L$ is a parameter), referred as \emph{target} strings, to use in the next step as markers to guide the factorization of the input reads.
The \emph{target} strings are extracted from each one of the factorizations of the considered $W$-long transcript substrings. Let $\fact_t(s) = \langle f_1, f_2, \ldots, f_n \rangle$ be one of such factorizations for substring $s$ of transcript $t$ and $\langle p_1, p_2 \ldots, p_n \rangle$ be such that $p_i$ is the starting position on $t$ of factor $f_i$.
Given a parameter $D$, we call \emph{target-list} the maximum subsequence $\langle p'_1, p'_2, \ldots, p'_q  \rangle$ of $\langle p_1, p_2 \ldots, p_n \rangle$ such
that $p'_1 = p_1$ and $p'_{i+1} - p'_{i} \geq D$ ($2 \leq i < q$). Each one of positions $p'_i$ on $t$ originates the \emph{target} string  $t[p'_i:p'_i+L-1]$.

\paragraph{Adaptive factorization of the long reads.}
The factorization of each input read $r$ is computed in an \emph{adaptive} manner, meaning that it is guided by the set $\cal F$ of the \emph{target} strings computed by the first step.
The read $r$ is partitioned into $n$ segments $w_1, w_2, \ldots, w_n$ such that $r = w_1 w_2 \ldots w_n$ and, for each $w_i$, there exists a \emph{target} string which is prefix of $w_i$. Then, the factorization of each $w_i$ is computed, obtaining $n$ fingerprints.
Intuitively, the \emph{target} strings aim at marking starting factor positions in the training factorizations and are used to obtain read segments that produce factorizations "aligned" with the training factorizations. In addition, the read segmentation concurs to reduce the factor length with respect to the factorization obtained by processing the whole read.

\paragraph{Classification of the long reads into chimeric and non-chimeric.}
For each long read $r$ and for each segment $w_i$, the $k$-fingers are extracted from the fingerprint of the segment and an ordered list $\langle \mathcal{K}_1, \mathcal{K}_2, \ldots, \mathcal{K}_q \rangle$ of $k$-fingers is produced, such that the $k$-fingers of $w_i$ come before the $k$-fingers of $w_j$ if $i < j$; moreover, the $k$-fingers of a given segment are listed from left to right in the segment.
At this point, the classifier assigns each $k$-finger to a gene, thus obtaining a list $List_G(w) = \langle g_1, g_2, \ldots, g_q \rangle$ such that $g_i$ is the gene to which $k$-finger $\mathcal{K}_i$ has been assigned.

On chimeric reads merging two genes, we expect that $List_G(w)$ contains a sublist for each gene where the gene is highly repeated.
To this aim, we measure the \emph{repetitiveness} of a given gene $g$ in $List_G(w)$ through the number of times two consecutive elements are equal to $g$ and, for each gene $g$, we compute the maximum sublist $\langle g_i, g_{i+1}, \ldots, g_j\rangle$, we call \emph{$g$-coverage}, such that $g_i = g_j = g$ and $g$ is the gene with the highest \emph{repetitiveness}.

\begin{figure}
	\centering
  \includegraphics[width=\linewidth]{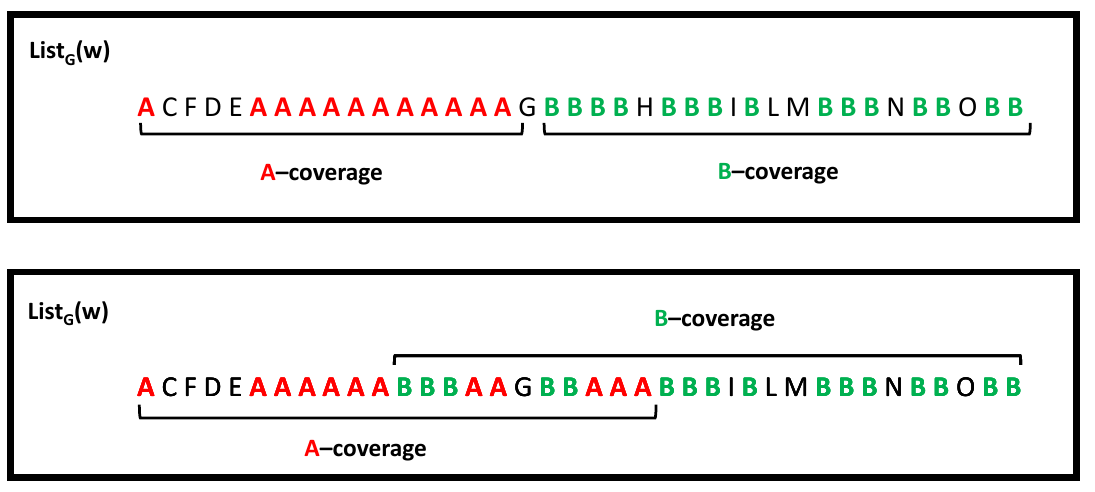}
  \caption{
    Two examples of a list $List_G(w)$ depicted as sequences of letters.
    Gene $A$ has repetitiveness 10 in the upper list and 8 in the lower list, while gene $B$ has repetitiveness 9 in the upper list and 9 in the lower list.
    Gene $A$ has a coverage of $0.42$ and $0.58$ in the upper and lower example (respectively), while gene $B$ has a coverage of
    $0.55$ and $0.71$ in the upper and lower example.
    The lower list presents an overlap between the two gene coverages and a
    percentage of common elements (with respect to the size of $List_G(w)$) equal to $0.29$. The fusion scores are $0.7$ and $1$
    for the upper and lower example, respectively.}
	\label{fig:coverage}
\end{figure}

For all pairs of genes $g_1$ and $g_2$, we compute a measure of how likely such genes are to be fused in the read.
Let $P_1$ ($P_2$, respectively) be the size of the $g_1$-coverage ($g_2$-coverage, resp.).
Let $P_{\cap}$ be the number of the common elements between the $g_1$-coverage and the $g_2$-coverage.
Then, we define the \emph{fusion score} for $g_1$ and $g_2$ as $score_w(g_1,g_2) = (P_1 + P_2 - P_{\cap}) / |List_G(w)|$. Observe that, $score_w(g_1,g_2)$ ranges from $0$ to $1$: $1$ means \textit{perfect coverage} of $List_G(w)$, while $0$ means that the two genes do not cover $List_G(w)$ at all.
Figure~\ref{fig:coverage} (where the genes are represented by letters) shows two examples of how the $g$-coverage of two fused genes may appear.

Hence, for each input long read $r$ and for each pair of genes $g_1$ and $g_2$, we compute the $g_1$-coverage and the $g_2$-coverage. If the $g_1$-coverage is not enclosed in the $g_2$-coverage (or vice versa), then the \emph{fusion score} $score_w(g_1, g_2)$ is obtained. Among all the pairs of considered genes, we select the pair giving the highest score and we label $r$ as chimeric if that score is at least a given threshold.

\subsubsection{Experimental results}

We tested our method on the PacBio reads used in~\cite{genfusdataset} (identity level of 95\%), and randomly selected $100$ genes from the set of considered genes.

The training was performed on the set of their $433$ protein-coding annotated transcripts
from the GENCODE annotation (version 22). The testing was performed on a balanced set composed of $2372$ chimeric reads and $2372$ non-chimeric reads.

We performed the classification of the long reads by using the following parameters: $W=300$ (length of the trasncript substrings in the training step), $D \in \{50, 100, 150, 200,$ $250, 300\}$ and $L \in \{8,16\}$ (for building the set $\cal F$ of the \emph{target} strings).

We used a minimum score threshold in $\{0.005, 0.01, \ldots, 0.5\}$.
The best results were obtained for $L = 8$, $D = 150$ and a minimum score $0.87$, achieving an \textit{accuracy} of $0.87$, a \textit{precision} of $0.81$, a \textit{recall} of $0.82$ and an \textit{F-score} of $0.81$.

\subsection{Discussion and perspectives}
\label{ssec:discussion}

The obtained results suggest that \texttt{lyn2vec} is effective in providing representations of sequencing reads to use in a Machine Learning framework.

The representation based on the basic notion of fingerprint (see Section~\ref{ssec:EQ1-task}) already provided results (precision 0.94 and recall 0.93) overcoming  those obtained using other well-known representation methods, \ie \texttt{BioVec}, \texttt{fastDNA} and \texttt{DNABERT}.

Later, in Section \ref{ssec:EQ2-task}, we further assessed our method by applying it on RNA-Seq reads simulated with Flux Simulator with different gene-expression levels and errors, showing that using a hybrid approach in which $k$-fingers were used in combination with a rule-based classifier, significantly increases performance, by reaching high precision 0.91 and recall 0.77, despite of a precision of 0.85 and a recall of 0.42  reached by directly classifying the whole fingerprints of such data.

Moreover, the use of superfingerprints
(Section~\ref{ssec:EQ3-task}) reduces the impact of collisions on the classification results and allows precision and recall to reach $0.95$ and $0.83$, respectively.

The $k$-finger representation produced by \texttt{lyn2vec} has been also tested in the gene fusion framework in order to detect chimeric reads, achieving a satisfactory performance: $0.87$ in accuracy, $0.81$ in precision, $0.82$ in recall and an F-score of $0.81$.

\section{Conclusion}
\label{sec:conclusion}
In this paper we propose \texttt{lyn2vec}, a novel feature embedding method exploiting the notion of fingerprint to represent sequencing reads. Differently from NLP-based feature embedding, our method relies on a theoretical investigation of combinatorial properties which guarantee to capture similarities among sequences.
The computational complexity of \texttt{lyn2vec} is linear in the number of represented sequences, \texttt{lyn2vec} does not require a training step and the size of the produced datasets is smaller with respect to those obtained with NLP-based sequence embedding methods.

We opted for testing the representations produced by \texttt{lyn2vec} in two frameworks: read-gene classification and chimeric read detection for inferring gene fusions. In particular, the chimeric read detection captures similarities when the strings (or reads) share only small portions (\ie a prefix or a suffix) with the two origin trasncripts; hence, it is an ideal context where assessing the soundness of \texttt{lyn2vec}.

We also performed an analysis to investigate the misclassification errors in the read-gene classification framework, pointing out that the performance of the classification by using $k$-fingers can be greatly improved by high values of $k$, which are possible when using long reads in place of short reads.

Undoubtedly, it is worth evaluating \texttt{lyn2vec} in other Machine Learning applications carrying out sequence analysis, such as clustering, structure prediction or even pattern discovery in order to develop practical applications on real data.

\section*{Funding}
This project has received funding from the European Union's Horizon 2020 research and innovation programme under the Marie Skłodowska-Curie grant agreement number [872539].

\bibliography{biblio-lyndon.bib}

\end{document}